\documentclass[twocolumn]{aastex631}
\usepackage{subfigure}
\usepackage[bookmarks=false]{hyperref}

\AtBeginEnvironment{figure}{\setcounter{subcaption}{0}}%
\AtBeginEnvironment{table}{\setcounter{subcaption}{0}}%

\makeatletter
\newcounter{subcaption}
\renewcommand{\thesubcaption}{(\alph{subcaption})}
\newcommand{\subcaption}[1]{
  \refstepcounter{subcaption}
  {\small\thesubcaption~#1}
}
\makeatother
\usepackage{graphicx}
\usepackage{natbib}
\usepackage[utf8]{inputenc}
\usepackage{CJK}
\usepackage{verbatim}
\submitjournal{ApJ}
\shortauthors{Le~Bourdais et al.}
\begin{document}

\begin{CJK*}{UTF8}{gbsn}
\title{Revisiting the Chemical Composition of WD~1145+017: Impact of Circumstellar Disks Contamination on Photospheric Abundances} 

\correspondingauthor{\'{E}rika Le Bourdais}
\email{erika.le.bourdais@umontreal.ca}

\author[0000-0002-3307-1062]{\'{E}rika Le Bourdais}

\author[0000-0003-4609-4500]{Patrick Dufour}
\affiliation{Trottier Institute for Research on Exoplanets and Department of Physics, Universit\'e de Montr\'eal, 1375 Ave. Th\'er\`ese-Lavoie-Roux Montr\'eal, QC H2V 0B3, Canada}

\author[0000-0002-8808-4282]{Siyi Xu (许\CJKfamily{bsmi}偲\CJKfamily{gbsn}艺)} 
\affiliation{Gemini Observatory/NSF's NOIRLab, 670 N A'ohoku Place, Hilo, HI 96720, USA}
 
\begin{abstract}
We performed a chemical analysis of the asteroid-bearing white dwarf WD~1145+017 using optical and ultraviolet spectroscopic data from 25 epochs between 2015 and 2023. We present an updated gas disk model with improved opacity calculations and temperature profiles to properly account for all circumstellar absorption features. Incorporating these changes into our models, we identified at least 10 elements in the disk, including a detection of circumstellar Na. We detected 16 elements in the photosphere, including new detections of P, Co and Cu. At 16 elements, WD~1145+017 ties GD~362 as one of the most polluted white dwarfs in terms of the number of elements detected. We find that both the disk and photosphere compositions align, to first order, with CI Chondrite. Our study underscores the importance of accounting for circumstellar absorption, as neglecting them leads to significant abundance errors. Additionally, the analysis of the disk's opacity highlighted a ultraviolet flux reduction due to a pseudo-continuum due to a optically thick component. This result may affect previous analyses of other polluted white dwarfs, suggesting a need for revisiting some studies.
\end{abstract}
 

\section{Introduction} \label{sec:intro}
\end{CJK*}
Since the discovery of the first metal polluted white dwarf over a century ago \citep{van_maanen_two_1917}, it has been estimated that at least 25\% to 50\% of all white dwarfs are contaminated by some traces of heavy elements \citep{koester_frequency_2014, zuckerman_metal_2003,wilson_unbiased_2019,manser_frequency_2024}. Since the settling times of these elements are much shorter than the cooling age of these stars \citep{paquette_diffusion_1986, koester_accretion_2009}, an external source, most likely accretion from tidally disrupted asteroids, planets, comets or planetesimals that once orbited the star, must have recently replenished the photosphere \citep[see][and references therein]{jura_extrasolar_2014}.\\

To date, 24 chemical species heavier than helium have been detected in the atmospheres of white dwarf stars \citep{klein_discovery_2021}, with a few dozen systems known to exhibit at least four different elements, according to the Montreal White Dwarf Database \citep{dufour_montreal_2016}. For the most polluted objects, the relative abundances of the elements generally resemble those of known rocky objects in the Solar System \citep[e.g.,][]{zuckerman_chemical_2007, jura_pollution_2008, xu_elemental_2014,swan_interpretation_2019,klein_chemical_2010,putirka_polluted_2021}. The detailed analysis of metal-polluted white dwarfs thus provides a unique opportunity to study in depth the bulk chemical composition of extrasolar bodies. Moreover, the structure and formation history of the accreted objects can potentially be inferred from detailed analysis. For example, an overabundance of oxygen may indicate accretion from a water-rich body, whereas the depletion of iron could suggest accretion of mantle-like material from a differentiated body \citep[see][for excellent reviews]{rogers_seven_2024, rogers_seven_2024-1,jura_extrasolar_2014}.\\

\begin{figure}[h]
    \centering
    \includegraphics[trim={3.7cm 0.79cm 2.4cm 0.7cm},clip,scale=0.085]{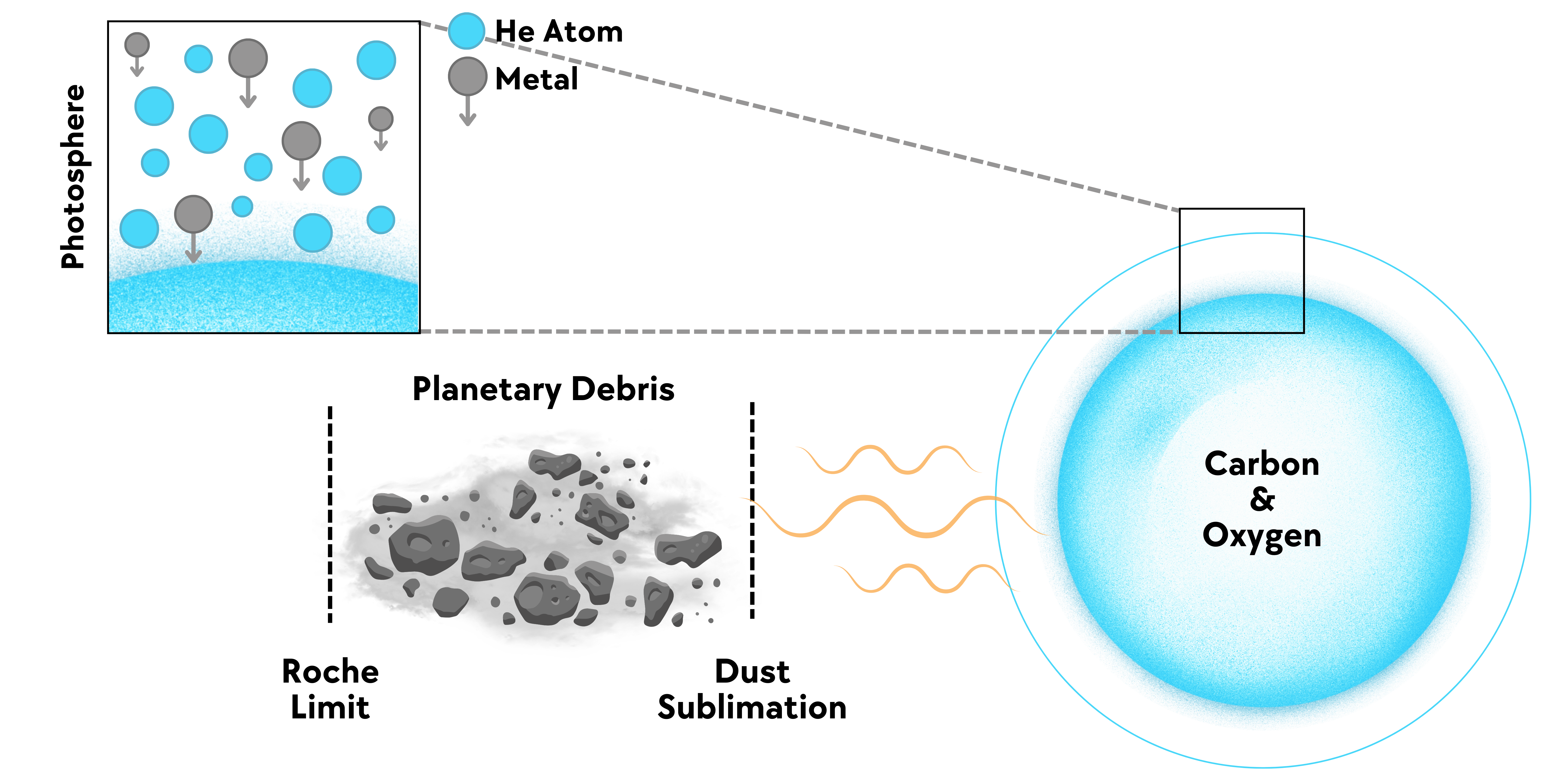}
    \caption{Illustration summarizing the mechanism leading to the pollution of white dwarfs. Adapted from \citet{jura_extrasolar_2014}.}
    \label{fig:cartoon}
\end{figure}

However, the derived element abundances are not necessarily straightforward to interpret. Several assumptions must be made, which necessitates a cautious approach. For example, assumptions about the accretion process can significantly impact the interpretation of abundances, potentially mimicking other types of material \citep{brouwers_asynchronous_2023-1, brouwers_asynchronous_2023, swan_planetesimals_2023}. The prevailing scenario for bringing material to the surface of a white dwarf, illustrated in Figure \ref{fig:cartoon}, involves several stages. First, an object is disturbed from its orbit and reaches the Roche limit of the white dwarf, where it disintegrates and forms a circumstellar disk. Then, due to Poynting-Robertson drag, a process caused by the star's radiation, the material gradually spirals inward and is slowly accreted into the star's photosphere \citep{debes_are_2002}.

\subsection{A primer on WD 1145+017}
The discovery of transiting debris around WD~1145+017 provided a new angle for studying the accretion process. This white dwarf showed multiple transits, reducing the star's brightness by up to 50\% with a period of approximately four and a half hours \citep{vanderburg_disintegrating_2015}. Not only are the transits of WD~1145+017 deep, but they also vary over time on scales ranging from a few hours to a few days \citep{rappaport_drifting_2016}. Interestingly enough, the transits disappeared around 2021 \citep{aungwerojwit_long-term_2024} while the spectrum still presents circumstellar gas absorption in recent spectroscopic data from 2023 (see Figure \ref{fig:duf12-2023-1}). These observations indicate the active disintegration and subsequent accretion of an extrasolar asteroid. Consequently, WD~1145+017 represents the first opportunity to observe the mechanism responsible for the metal enrichment of white dwarf stars in real time.\\

Following the discovery of transits, \citet{xu_evidence_2016} reported numerous circumstellar absorption features with linewidths of approximately 300~km/s from several elements. These features are most likely the result of absorption from high velocity gas streams produced by collisions among the actively disintegrating parts of the asteroid. These circumstellar features blend with photospheric absorption features, complicating significantly the chemical analysis of this object. Subsequently, the circumstellar absorption line profiles showed complete reversals in velocity, shifting from strongly redshifted to strongly blueshifted, indicating the presence of a precessing eccentric disk \citep{cauley_evidence_2018}.\\

Inspired by the work of \citet{cauley_evidence_2018}, a simple gas disk model was later developed by \citet[][hereafter FA20]{fortin-archambault_modeling_2020}. This model uses 14 eccentric precessing rings to replicate the circumstellar spectroscopic features in both the optical and UV spectral regions. The model employed the radiative transfer code \textsc{tlusty205} and \textsc{synspec51} \citep{hubeny_brief_2017}, assuming a disk temperature of 6000 K and an inner midplane density of $6 \times 10^{-6} {\rm~g/cm}^3$. Although this model was relatively successful at reproducing many absorption features of Fe, some regions could only be reproduced by introducing a linear vertical temperature gradient, which lacked a strong physical basis. While it significantly improved the Ti circumstellar features, some regions were still not well fitted.\\

More recently, \citet{budaj_wd_2022} (hereafter B22) published a different model for the disk surrounding WD~1145+017, which provides a more accurate representation of the physical properties of gas and dust disks compared to the FA20 model. This model consists of two disks, one for gas and one for dust, and employs a more physically realistic radial temperature profile. The gas disk has an inner temperature of 5700~K and an inner midplane density of $3.3\times 10^{-12}{\rm ~g/cm}^3$, which is six orders of magnitude less than in FA20. This discrepancy prompted an investigation into the reason for such a significant difference and is discussed in section \ref{sec:res_disk-param}. Although the B22 disk model is more physically accurate in many aspects, it does not reproduce the features as well as the FA20 model. Additionally, since the authors tested their model on only one line, namely the \ion{Fe}{2} 5316~\AA\ line, it is unclear how it would perform for other elements in the UV and optical spectra.\\

Since the publication of the FA20 toy model, many new epochs of observation from both ground-based and space-based telescopes have become available. In this context, it was deemed an opportune moment to correct and improve upon the FA20 model and analyze all the spectroscopic data available. With this paper, our goal is to use our improved model to accurately represent all circumstellar features from UV to optical wavelengths, thereby allowing for a more precise determination of the chemical composition of the photosphere.\\

A description of the spectroscopic data used in this study is presented in Section \ref{sec:obs}. Our theoretical framework and methods for both the disk and the stellar photosphere are detailed in Section \ref{sec:methods}. Section \ref{sec:results} presents the results of our modeling of the disk and photosphere. Section \ref{sec:discussion} discusses the limitations of our disk model as well as the potential impact of our results on abundances determination of other polluted white dwarfs. Finally, Section \ref{sec:conclusion} summarizes our findings and conclusions.\\

\section{Observations} \label{sec:obs}

Part of the data and their respective reduction procedure used for this work are described in \citet{xu_shallow_2019} and \citet{fortin-archambault_modeling_2020}. This dataset comprises 17 epochs of intermediate and high-resolution spectra in the optical and UV regions from the Keck I telescope, VLT, and HST. In addition to these data, we also use new observations from HST-COS (3 epochs from 2018 and 2019) and the newly commissioned Gemini High-resolution Optical Spectrograph (GHOST) instrument on the Gemini South telescope (data taken on May 10, 2023). \object{WD~1145+017} was observed with GHOST (\citealp{mcconnachie_science_2024}) as part of the system verification process. The observations were conducted under the program GN-2023A-SV-103 on May 10, 2023 (UT). The standard resolution mode was used, providing a resolving power of 56,000. A binning of 1 by 2 was adopted. The observing conditions were decent, with seeing around 0{\farcs}8 and thin cirrus. Two 30-minute exposures were obtained, with a flux standard LTT~7379 observed using the same setup. Data reduction, including flux calibration, was performed using version 1.0.0 of the GHOST data reduction pipeline with DRAGONS \citep{labrie_dragons-quick_2023}. The final spectrum has a wavelength coverage of 3500--10,000~{\AA}.\\

With this new set of data, the observed time frame now comprises a total of 25 epochs from 2015 to 2023. The observations used for this paper are listed in Table \ref{tab:WD-obs}.

\begin{table*}
	\centering

	\begin{tabular}{lllll} 
		\hline
  \hline
		Date & Telescope & Instrument & Resolution & Spectral range~{(\AA)}\\
		\hline	
		2015 April 11& Keck I & HIRESb & 35,800 & 3050--5940\\
        2015 April 25& Keck II & ESI & 13,750 & 3900--10,900\\
        2016 February 3& Keck I & HIRESr & 35,800 & 4690--9140\\
        2016 March 3& Keck I & HIRESb & 35,800 & 2700--5660\\
        2016 March 28& Keck II & ESI & 13,750 & 3900--10,900\\
        2016 March 28& HST & COS/FUV & 18,000 & 1125--1440\\
        2016 March 29& VLT & X-Shooter (UVB \& VIS)& 6200 \& 7400 & 3100--10,000\\
        2016 April 1& Keck I & HIRESb & 35,800 & 3050--5940\\
        2016 November 18& Keck II & ESI & 13,750 & 3900--10,900\\
        2016 November 19& Keck II & ESI & 13,750 & 3900--10,900\\
        2016 November 26& Keck I & HIRESr & 35,800 & 4690--9140\\
        2016 December 22& Keck I & HIRESr & 35,800 & 4880--9380\\
        2017 February 17$^*$ & HST & COS/FUV & 18,000 & 1125--1440\\
        2017 February 18$^*$ & HST & COS/FUV & 18,000 & 1125--1440\\
        2017 March 6& Keck II & ESI & 13,750 & 3900--10,900\\
        2017 March 7& Keck II & ESI & 13,750 & 3900--10,900\\
        2017 April 17& Keck II & ESI & 13,750 & 3900--10,900\\
        2017 June 6$^*$&HST&COS/FUV& 18,000 & 1125--1440\\
        2017 June 6$^*$$^{\dagger}$&HST&COS/FUV& 18,000 & 1125--1440\\
        2017 April 30$^*$&HST&COS/FUV& 18,000 & 1125--1440\\
        2018 January 1&Keck I&HIRESb& 35,800 & 3050--5940\\
        2018 April 24&Keck I&HIRESb& 35,800 & 3050--5940\\
        2018 April 30$^*$&HST&COS/FUV& 18,000 & 1125--1440\\
        2018 May 1$^*$&HST&COS/FUV& 18,000 & 1125--1440\\
        2018 May 18&Keck I& HIRESb& 35,800 & 3050--5940\\
        2018 June 14$^*$&HST&COS/FUV& 18,000 & 1125--1440\\
        2023 May 10$^*$&Gemini South&GHOST& 56,000 & 3500--10,000\\
		\hline
        \multicolumn{5}{l}{$^*$New data published in this paper \hspace{1cm} $^\dag$Observed in transit}\\
	\end{tabular}
      \caption{Observations of WD~1145+017 used for this paper.}
	\label{tab:WD-obs}
\end{table*}

\section{Methods and models}\label{sec:methods}
\subsection{Photospheric parameters}\label{sec:photometry}

The effective temperature of WD~1145+017 was first estimated by \citet{vanderburg_disintegrating_2015} to be $T_{\rm eff}$ = 15,900~K from a photometric fit, assuming a $\log g = 8.0$ and an approximate chemical composition. A new photometric fit was later presented in FA20, using the newly measured Gaia trigonometric parallax and assuming a more realistic chemical composition following the method of \citet{coutu_analysis_2019}. They obtained a much lower effective temperature of $T_{\rm eff}$ = 14,500~K, but our subsequent analysis of their fit discovered that the interstellar reddening had not been incorporated correctly. While this error was of no consequence for the circumstellar disk, the lower effective temperature possibly affects the derived abundances. Fortunately, the metal-to-metal ratios, and thus the interpretation of the polluting object's composition, are much less affected by small changes in effective temperature.\\

\begin{figure}[h]
    \centering
    \includegraphics[trim={0.2cm 0.1cm 1.5cm 1.5cm},clip,scale=0.53]{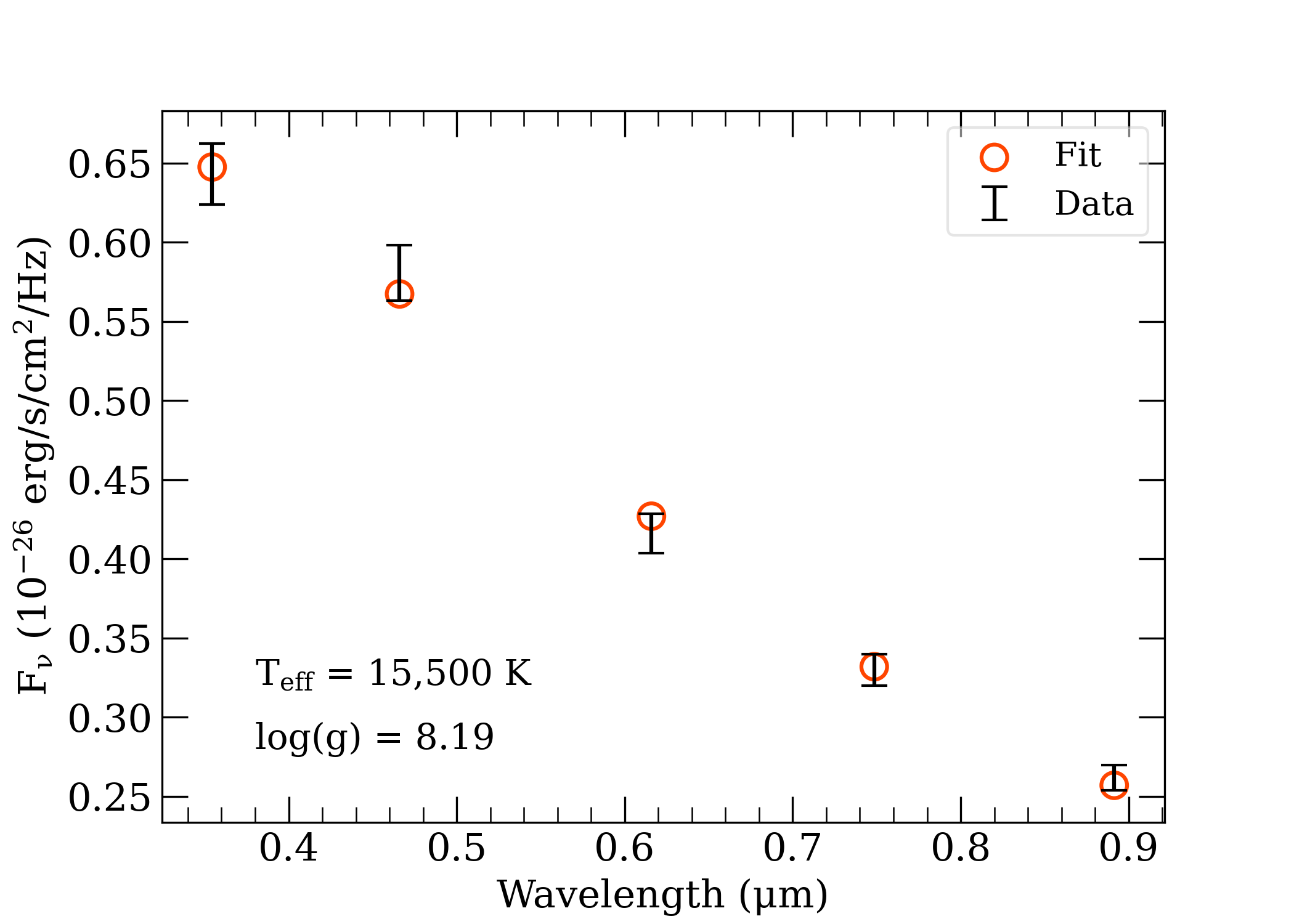}
    \caption{Photometric measurements in SDSS \textit{ugriz} bands compared to our best-fit model for WD~1145+017.}
    \label{fig:photometry}
\end{figure}

We thus performed a new photometric fit on the SDSS \textit{ugriz} spectral bands (see Figure \ref{fig:photometry}) using the method of \citet{coutu_analysis_2019}, using Stilism \citep{capitanio_three-dimensional_2017} extinction on the line of sight to WD~1145+017 and chemical abundances close to our final findings. We arrived at $T_{\rm eff} = 15,500 \pm 500$~K and $\log g = 8.19 \pm 0.03$, which we adopt for the rest of this study.\\

\subsection{Gas Disk Model and Photospheric Abundances}\label{sec:disk}
Our aim in this study is to reproduce the various absorption features for a given epoch with a simple configuration, rather than to make future predictions about the behavior of the disk. Still, a few modifications to the FA20 disk model have been made for this work. First, we now utilize the more flexible opacity tables feature in the latest release of \textsc{synspec54} \citep{hubeny_tlusty_2021}, which replaces the previously used iron curtain mode that had erroneous units (see below). We also implemented the same radial temperature profile used by \citet{budaj_wd_2022}, which is a typical profile for a gaseous accretion disk \citep{pringle_accretion_1981}. We find that the use of a vertical temperature profile, as used by FA20 to help reproduce the steep Ti circumstellar features, was no longer necessary, and thus removed it as there was no physical justification for its use.\\

We maintain the same disk configuration as in \citet{fortin-archambault_modeling_2020}, consisting of 14 eccentric confocal rings evenly spaced radially, as well as the precessing scenario described, despite its inconsistency with predictions from general relativity. Finally, we increased the grid size to a $50\times 50$ coverage in the plane perpendicular to the line-of-sight, allowing for better resolution of the midplane region which contributes most of the absorption. This improvement is particularly significant in the UV region since there are more spectral lines in this spectral region (see Section \ref{sec:UV}).\\

In light of these upgrades, we reexplored the parameter space for new values of temperatures and densities, ranging from 4000 to 9000 K and from $1\times10^{-12}$ to $1\times10^{-11}$~g/cm$^3$, respectively. The process to determine the temperature, density, and abundance was iterative due to the intrinsic degeneracy of all parameters and the blending with photospheric spectral lines. Starting with the abundances from a typical CI Chondrite composition \citep{lodders_solar_2003} and including the first 30 elements of the periodic table (H to Zn), we first identified the range of densities and temperatures that somewhat matched the observations. We then manually adjusted the abundances as needed. This process was repeated until we found the best fit for most of the optical circumstellar features and the configuration closest in time to the HST observation date for the UV data.\\

Once the ideal configuration of the disk was found for each available epoch of observations, we performed a detailed analysis of the chemical composition of WD~1145+017. For each epoch, we kept the same physical parameters in the disk. Using a thermodynamic structure calculated assuming the effective temperature and $\log g$ presented in Section \ref{sec:photometry}, we then calculate grids of synthetic spectra for each element. We proceed by iteratively fitting the various spectral features following the method described in \citet{dufour_detailed_2012}, with the important difference that we also consider the contribution of the circumstellar disk by combining the disk model with our synthetic spectra. This consideration is crucial due to the severe blending between these two contributions (see Figures \ref{fig:20160401} and \ref{fig:20160328}). The technical details about the integration of the disk model to the white dwarf's synthetic spectrum is described in length in \citet{fortin-archambault_modeling_2020}. Final abundances were determined by averaging the abundances derived from all fitted epochs. Given that the sinking timescales of metals in helium-dominated white dwarfs are on the order of $10^5$ years, we can reasonably assume that the abundances remain constant over the timespan covered by the data. The final uncertainty accounts for both the variations across individual epochs and the fits performed using white dwarf models with temperatures varied by $\pm 500 ~\rm K$, with uncertainties calculated using the root mean square method.
 
\section{Results}\label{sec:results}
\subsection{Disk parameters}\label{sec:res_disk-param}
\begin{figure}[h!]
    \centering
    \includegraphics[trim={0.37cm 0.06cm 1.64cm 1.05cm},clip,width=\linewidth]{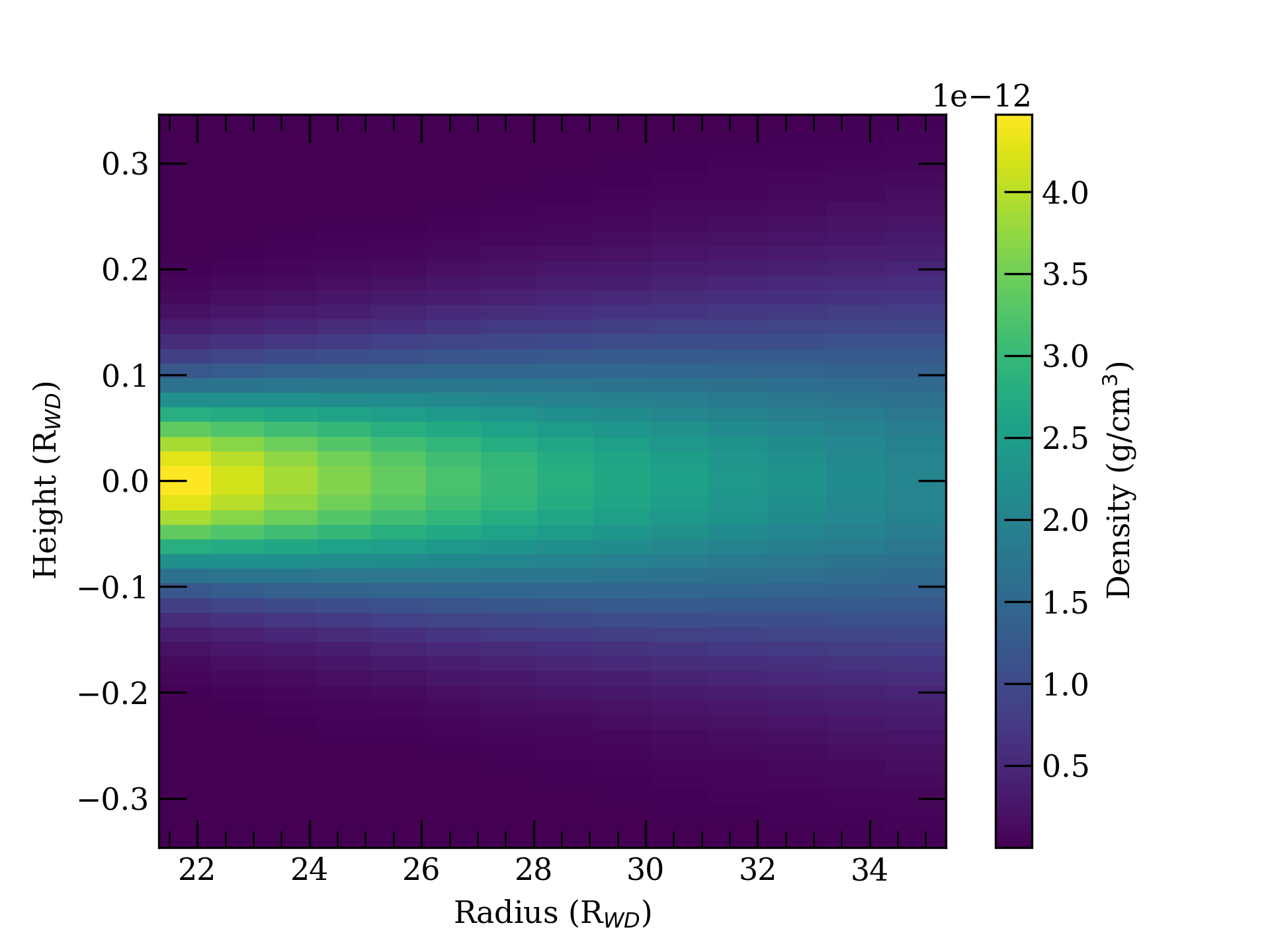}
    \caption{Side view map of the density distribution in the disk model.}
    \label{fig:density_map}
\end{figure}

We find that a midplane disk temperature of $7000 \pm 500$~K and a density of $4.5 \pm 2.1 \times 10^{-12} {~\rm g/cm}^3$ provide an excellent representation of most absorption features. Integrating our density profile, illustrated in Figure \ref{fig:density_map}, over the volume of the disk, we estimate a total mass of $6.8 \times 10^{17} {~\rm g}$, which interestingly represents only a tiny fraction of the mass of Ceres ($6.4 \times 10^{-7}M_{\rm Ceres}$).\\

Notably, this updated density value is much more consistent with what was expected in \citet{budaj_wd_2022} than the $6\times 10^{-6}{~\rm g/cm^3}$ obtained by \citet{fortin-archambault_modeling_2020}. Upon investigation, it was realized that the advertised units of opacity for iron curtain mode in the \textsc{synspec} user guide were erroneous (I. Hubeny, private communication). Consequently, the FA20 study had to increase the density to ensure that the product of density and opacity was of the correct order of magnitude to reproduce the absorption features. Since opacity calculations also depend on density, this compensation largely account, for the difficulty of FA20 in simultaneously fitting all features with the same temperature profile.\\

\begin{figure*}[h!]
    \centering
    \includegraphics[trim={3.20cm 3.6cm 3.4cm 5.9cm},clip,scale=0.54]{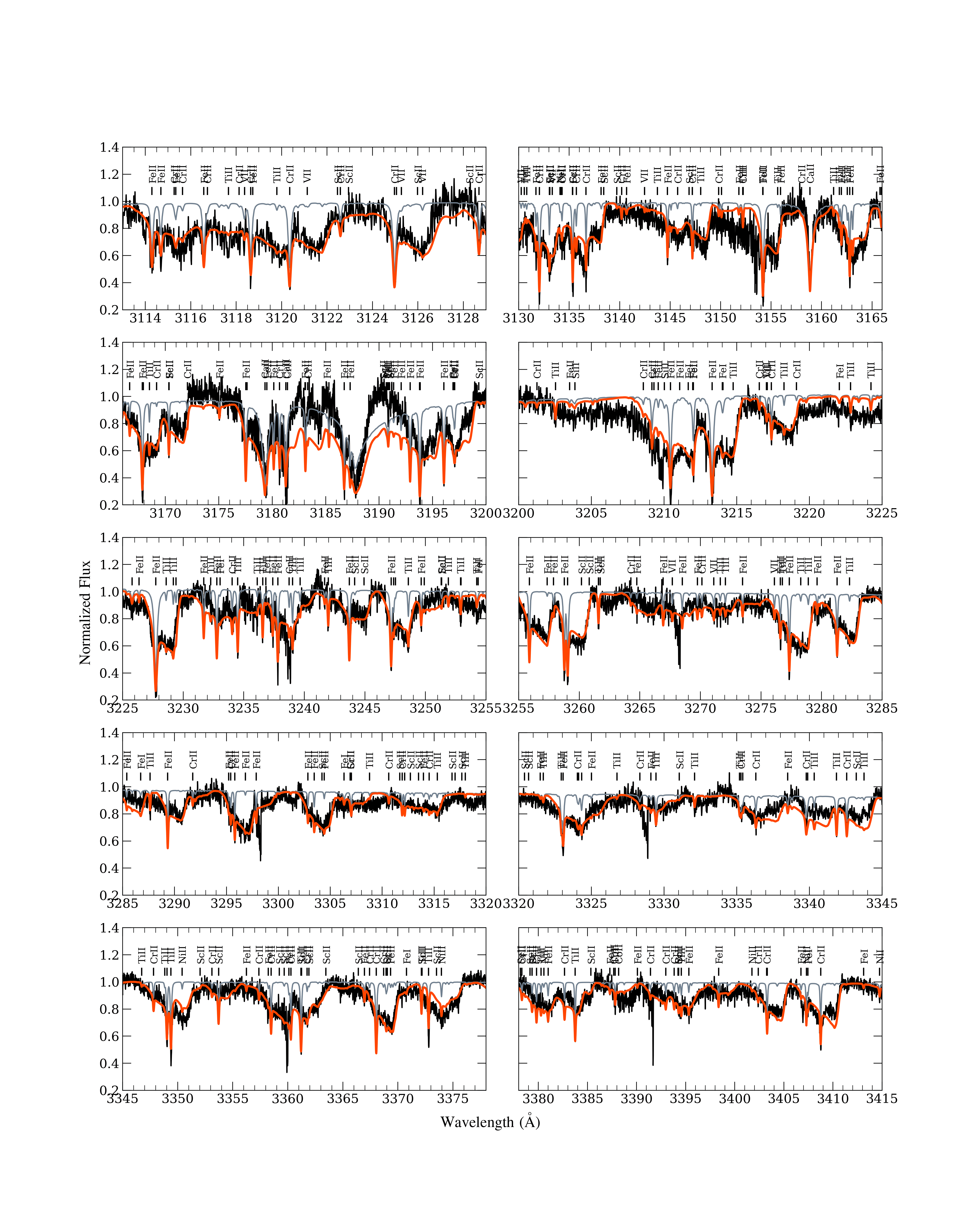}
    \caption{Display of our final solution over selected regions of the 2016 April 1 HIRES spectrum. The contribution from the photosphere alone is in grey while the combined photospheric and circumstellar disk contributions is in orange. The models and data used to produce the figures in this paper are available upon request.}
    \label{fig:20160401}
    
\end{figure*}

A comparison of our new model with observations from April 1, 2016, shown in Figure \ref{fig:20160401}, demonstrates that our new model can provide an excellent representation of the circumstellar contribution at almost all optical wavelengths. It is particularly important to include these contributions to correctly determine photospheric abundances, especially since it is rather difficult to precisely identify the continuum level if only fitting a small region centered on an individual photospheric line. In \citet{xu_evidence_2016}, abundance determinations were made using a set of lines believed to be uncontaminated by circumstellar absorption. In retrospect, practically all the features were contaminated to some degree, and the derived abundances were consequently affected.

\begin{figure*}[!h]
    \centering
    \includegraphics[trim={3.20cm 3.9cm 3.4cm 5.9cm},clip,scale=0.54]{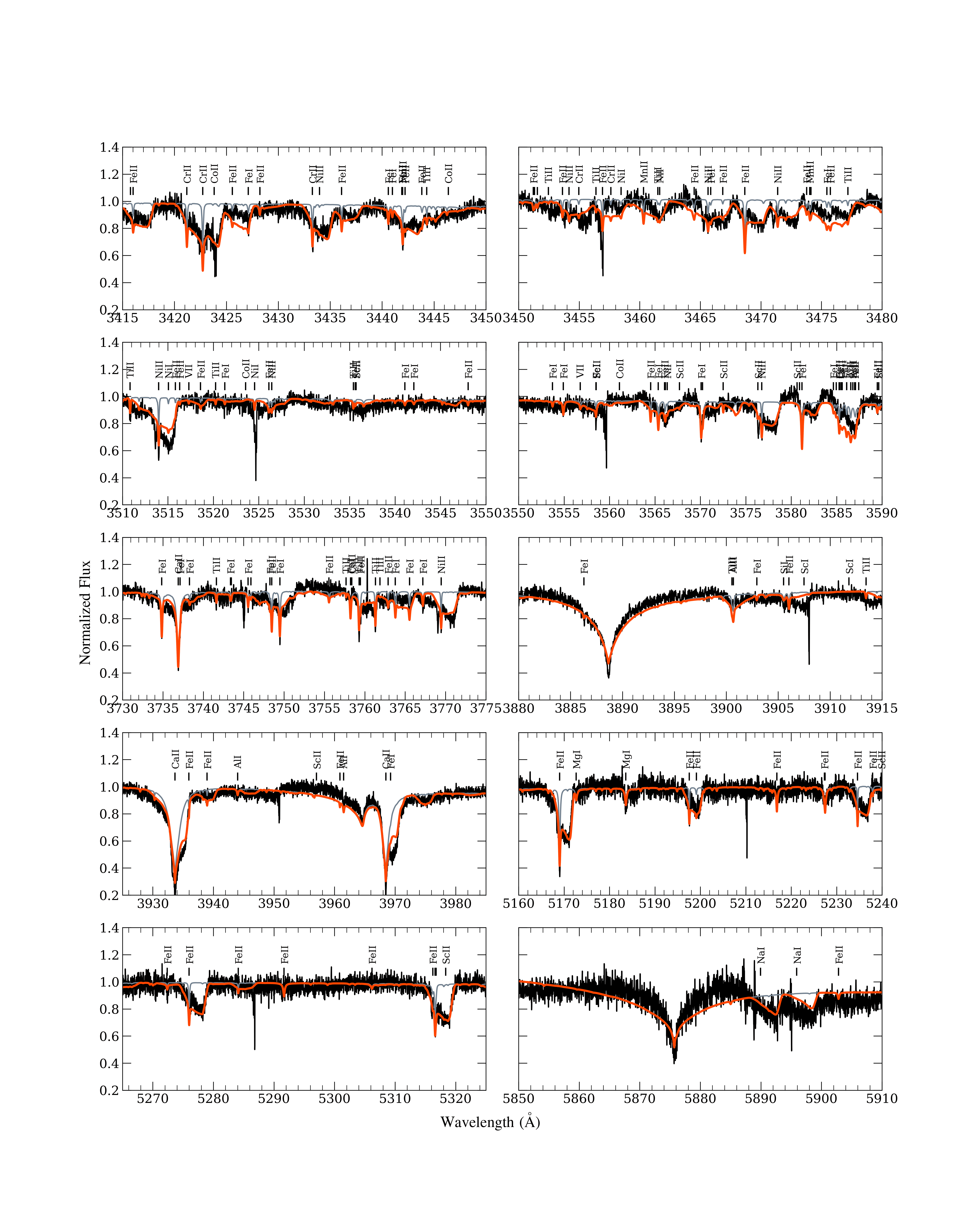}
    \subcaption{Figure \ref{fig:20160401}, continued.}
    \label{fig:20160401a}
\end{figure*}

\subsection{Chemical composition of the disk}\label{sec:disk_abn}
We identify that at least 10 distinct elements are contributing to the disk's absorption features (Fe, Al, Mg, Ni, Ti, O, Cr, Na, C, Ca), and there are hints of absorption from a few more elements that we can't confidently identify due to the very high velocity dispersion (200--300 km/s) of the faint features they produce. We also note the presence of large velocity-broadened Na features near 5889.95 and 5895.93~\AA\ (see panel 2 of Figure \ref{fig:duf12_20160203-1}). Since WD~1145+017 is too hot to show those lines in the photosphere, they must be circumstellar. However, a much lower disk temperature (around 4500 K, with a corresponding midplane density of $6\times 10^{-12} {\rm ~g/cm}^3$) is needed to reproduce those features. It is possible that this discrepancy in the disk parameters arises either from limitations in our assumed disk structure or from a concentration of sodium at outer radii. At the moment, our model assumes that everything is homogeneously mixed in the disk.

\begin{figure*}[ht!]
    \centering
    \includegraphics[trim={0.8cm 0.7cm 2.5cm 1.5cm},clip,scale=0.7]{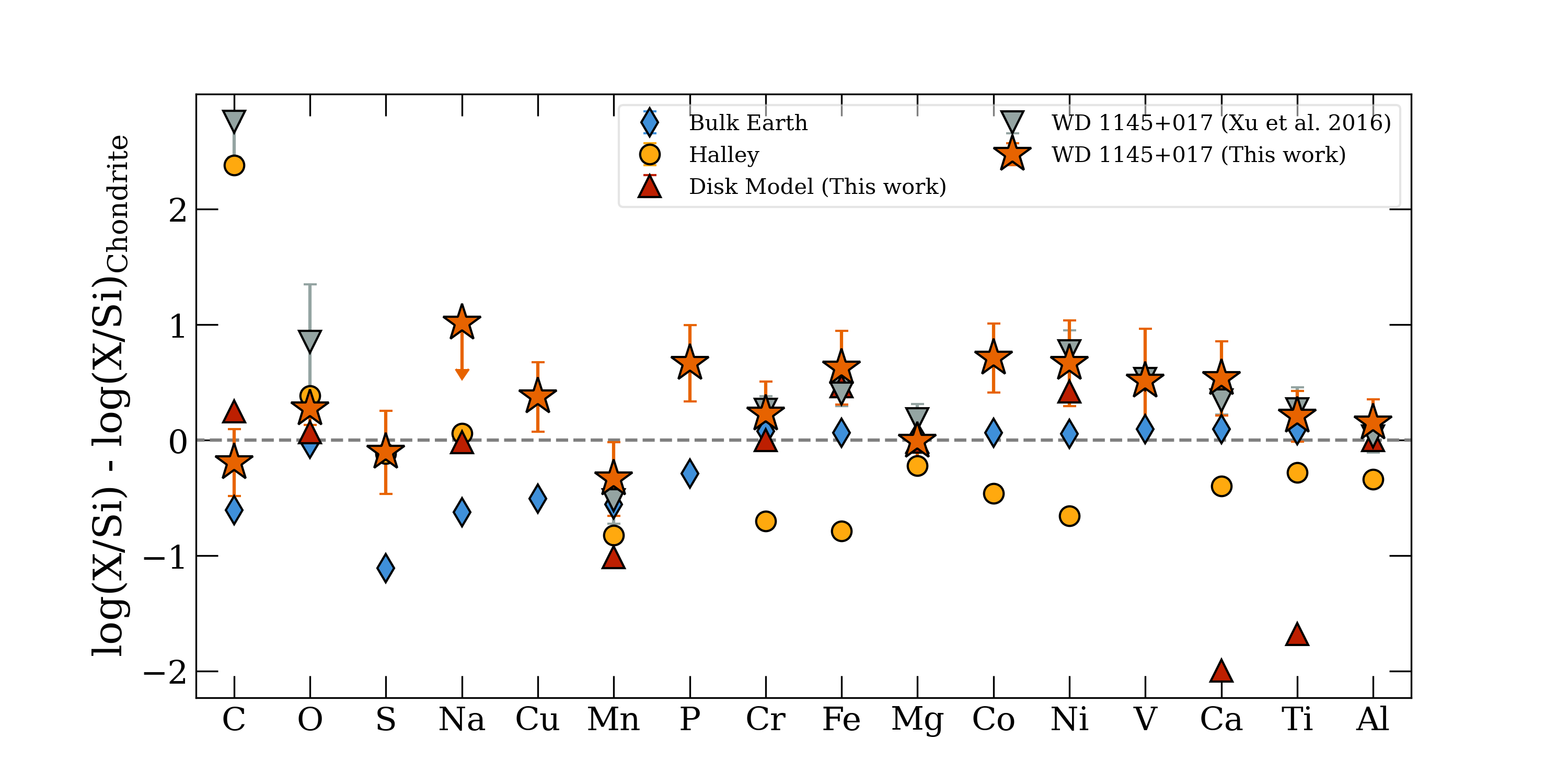}
    \caption{Photospheric $\log {\rm X/Si}$ relative to its ratio in CI Chondrite  \citep[dashed line,][]{lodders_solar_2003}, assuming steady state, in comparison to Bulk Earth \citep{mcdonough_compositional_2003} and comet Halley \citep{jessberger_aspects_1988}, as well as the previous abundances for WD~1145+017 from \citet{xu_evidence_2016}. Elements are ordered by condensation temperature. The abundances for the disk model are also shown for the elements with a clear circumstellar detection.}
    \label{fig:abn_ratios}
\end{figure*} 

\subsection{Photospheric Chemical Abundances}\label{sec:abn}
The derived photospheric abundances, averaged over all available epochs and from both optical and UV spectra, are presented in Table \ref{tab:abn} for our nominal atmospheric parameters of $T_{\rm eff}$ = 15,500~K and $\log g$ = 8.19. Examples of the fits we achieved for a few other epochs are presented in Figures \ref{fig:20160401} and \ref{fig:20160328}. A few other epochs are also presented in Appendix \ref{apx:fit_epochs}.\\

\begin{table}[!h]
    \centering
    \caption{Calculated photospheric abundances for WD~1145+017 in both the optical and UV spectral regions. Abundances from \citet{xu_evidence_2016}, which were obtained from optical spectra only, are also presented for comparison.}
    \label{tab:abn}
    \begin{tabular}{lccc}
    \hline
    \hline
         &\multicolumn{2}{c}{This work} &\citet{xu_evidence_2016}\\
        \textbf{$T_{\rm eff} (K)$}& \multicolumn{2}{c}{$15,500\pm500$}&$15,900\pm500$\\
        $\log g$&\multicolumn{2}{c}{$8.19\pm0.03$}&8.0\\
  \hline
  Element&\multicolumn{3}{c}{\textbf{log(Z/He)}}\\
  &Optical&UV &\citet{xu_evidence_2016}\\
  \hline
  H&$-4.83 \pm 0.14$&---&$-4.7 \pm 0.10$\\
  C&---&$-7.39 \pm 0.29$&$<-4.3$\\
  O&$-4.95 \pm 0.14$&$-5.43 \pm 0.35$&$-4.3 \pm 0.50$\\
  Na*&$<-6.0$&---&---\\
  Mg&$-5.75 \pm 0.28$&$-5.98 \pm 0.26$&$-5.49 \pm 0.13$\\
  Al&$-6.91 \pm 0.26$&$-6.71 \pm 0.16$&$-6.74 \pm 0.14$\\
  Si&$-5.79 \pm 0.14$&$-5.69 \pm 0.28$&$-5.69 \pm 0.09$\\
  P*&---&$-7.30 \pm 0.33$&---\\
  S&---&$-6.33 \pm 0.36$&---\\
  Ca&$-6.64 \pm 0.32$&---&$-6.57 \pm 0.12$\\
  Ti&$-8.39 \pm 0.22$&---&$-8.04 \pm 0.19$\\
  V&$-9.01 \pm 0.40$&---&$-8.7$:\\
  Cr&$-7.62 \pm 0.28$&---&$-7.31 \pm 0.12$\\
  Mn&$-8.34 \pm 0.32$&---&$-8.25 \pm 0.20$\\
  Fe&$-5.39 \pm 0.32$&$-5.46 \pm 0.31$ &$-5.35 \pm 0.11$\\
  Co*$^\dag$&---&$\sim -7.60 \pm 0.30$&---\\
  Ni&$-6.61 \pm 0.37$&$-6.79 \pm 0.32$&$-6.24 \pm 0.18$\\
  Cu*&---&$\sim -9.00 \pm 0.30$&---\\
  \hline
  \multicolumn{4}{l}{\footnotesize $^*$New element detected  $^\dag$Tentative detection}\\
\end{tabular}
\end{table}

 A total of 16 distinct elements, including three new additions from this study (P, Co and Cu), are detected in the photosphere of WD~1145+017, tying GD~362 as the most polluted white dwarf in terms of the number of elements detected. The inclusion of the circumstellar absorption component in our fitting procedure has a non-negligible impact, with differences of up to 0.3 dex found for certain elements when compared to \citet{xu_evidence_2016}. In Figure \ref{fig:abn_ratios}, we show the element abundance ratios relative to Si and normalized to CI Chondrite. To the first order, our results are compatible with CI Chondrite composition, with some differences for a few elements, notably P, Co, and Ni. However, the abundance of these elements is derived from only a few weak lines in regions highly contaminated by circumstellar features. As such, any misprediction of the "pseudo-continuum" absorption from the disk would translate into a significant change in the derived abundance. Therefore, the formal statistical error reported in Table \ref{tab:abn} could be underestimated.\\

\begin{figure}[h]
\includegraphics[width=1.01\linewidth,trim={0.8cm 0.1cm 1.4cm 1.8cm},clip]{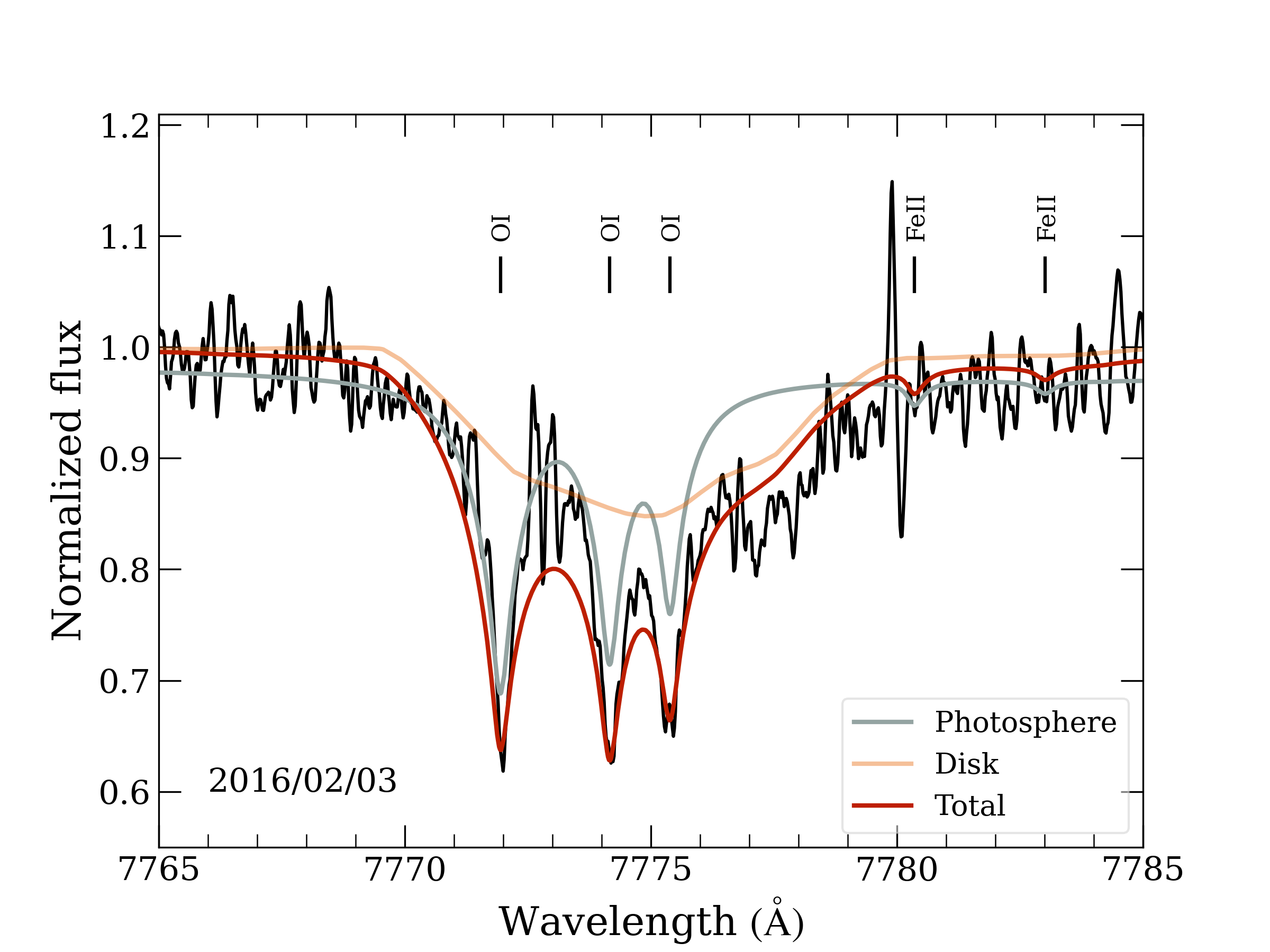} 
\caption{Disk and photospheric contribution to the spectrum in the \ion{O}{1} 7775 {\AA} triplet region for the 2016 February 3 HIRES data.}
\label{fig:oxygene}
\end{figure}

In previous photospheric analyses presented by \citet{xu_evidence_2016} and FA20, a large discrepancy between the oxygen abundance obtained from the \ion{O}{1} 7775{\,\AA} triplet and \ion{O}{1} 8446{\,\AA} line was found. In the FA20 model, the \ion{O}{1} 7775{\,\AA} triplet was not found to have a circumstellar contamination. Figure \ref{fig:oxygene} shows that this is no longer the case with our new model and with the inclusion of the disk's contamination in the fitting procedure, the derived abundance from both set of lines are now in excellent agreement. \citet{xu_evidence_2016} mentioned in their analysis that their oxygen abundance was particularly elevated to the point of being one of the most oxygen-polluted extrasolar planetesimals ever observed. Our disk-corrected abundance brings back the oxygen abundance toward more normal values. We note, however, that there is still a $\sim 0.5$~dex difference between the optical and UV abundances. These descrepancies have been found on occasions before for Fe and Si \citep[e.g.,][]{melis_does_2017,rogers_seven_2024,jura_two_2012,xu_compositions_2019,gansicke_chemical_2012} so it is possible that whatever the cause of this discrepancy for those elements may be in play as well for oxygen. One must keep in mind, however, that the uncertainties in the UV oxygen determination could be underestimated, given it is based essentially on only one line, namely the 1152~\AA\ line (nighttime data were seldom available for the 1300--1310~\AA\ region). Again, any misprediction of the circumstellar absorption in that region would consequently affect the derived photospheric abundance.

\section{Discussion}\label{sec:discussion}
While the improvements to the FA20 model eliminated the need for a vertical temperature profile to reproduce the Ti features, we did have to reduce Ti's abundance relative to CI Chondrite. Ti is particularly sensitive to the combination of temperatures and densities assumed in the disk. As shown in Figure \ref{fig:abn_ratios}, Ca was also reduced from CI Chondrite in the disk for similar reasons. Also, we noticed that while the model could reproduce the circumstellar features for the \ion{Ca}{2} H and K lines with the adjusted abundance, the model could not reproduce those around the 8600{\,\AA} \ion{Ca}{2} triplet using our actual disk parameters. We also found that we could not always perfectly reproduce all other features simultaneously with the same parameters, and that the best combination of temperature and density varied slightly from one element to another. Ultimately, since Fe is the principal contributor to the circumstellar features, we adopted parameters that best reproduced the \ion{Fe}{1} and \ion{Fe}{2} lines simultaneously.\\

It is interesting to note that no circumstellar Si features are detected, and its abundance in the disk had to be reduced for these features to remain absent. The circumstellar features that align with photospheric lines are attributed to Fe rather than Si. Given that Si is normally one of the four major components of rocky material, this absence is particularly surprising. However, it would be premature at this point to draw definitive conclusions about the chemical composition of this asteroid. Given the degeneracy between the physical parameters of the disk, it is also possible that acceptable modifications to the temperature structure, density, and abundances may lead to a solution that does not produce Si and Ca features while still maintaining a composition that is not too far from typical rocky objects. We leave the exploration of such configurations to future studies.\\

Finally, we remind the readers that while our model reproduces most absorption features well, it remains a relatively simple toy model. Any conclusions regarding the exact physical parameters of the circumstellar disk should be approached with caution. There is some degeneracy between temperature, density, and abundances in the disk model. The degeneracy lies in the fact that a small change in temperature can be offset by corresponding adjustments in density and the various abundances, leading to similar observational outcomes.
Even if the temperature and density are within realistic values, a more physical model is needed to confidently determine the disk's chemical composition, density, and temperature. We emphasize that for the purpose of this study, the exact parameters are of secondary importance. Our goal with this model is to mimic the circumstellar features well enough to extract abundances from photospheric features that are severely blended with circumstellar absorption.

\subsection{Sources of opacity in the UV}\label{sec:UV}
Now equipped with a model that reproduces very well most of the circumstellar optical features for each epoch, we can make predictions about the disk's contribution in the UV part of the electromagnetic spectrum. Figure \ref{fig:20160328} shows an example of our model with (orange line) and without (grey line) the contribution from the disk. Since the density of lines in the UV is much higher than in the optical, the numerous velocity-broadened lines almost form a constant pseudo-continuum that can absorb as much as 30\% of the flux at certain wavelengths. \\

It is thus even more crucial than in the optical to account for this flux reduction to accurately determine chemical abundances from UV spectra. While it can be argued that the exact level of additional opacity is not known with certainty, given that the UV model is based on the disk paremeters determined using the optical spectra, we have reason to believe that our model performs well in the UV as well. Our high confidence is based on the fact that at certain wavelengths, there are fewer lines contributing to the total opacity of the disk. In fact, in the midplane section, the disk is practically optically thick at certain wavelengths and gradually becomes thin as we move away from the middle. However, wavelengths with lower opacity offer windows through which the flux from the star can pass, making it appear almost as if there are emission lines.\\

\begin{figure}[ht!]
    \centering
    \includegraphics[trim={2.02cm 1.15cm 3.01cm 3.04cm},clip,scale=0.335]{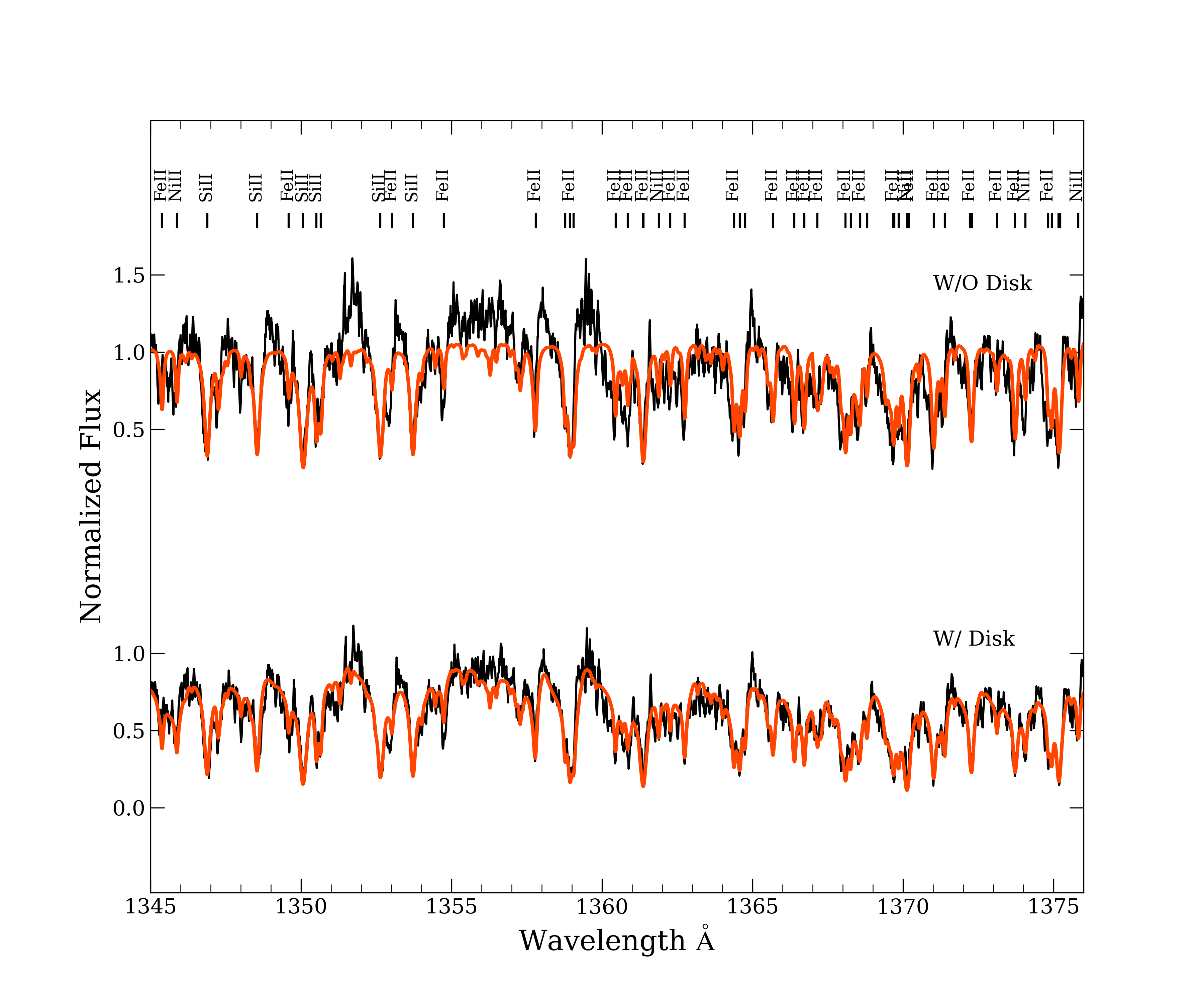}
    \caption{Example of a region with emission-like features for the 2017 February 17 HST COS data. The top spectrum shows our best fit without including the disk contribution, while the bottom spectrum shows the fit with the disk contribution included.}
    \label{fig:emissionUV01}
\end{figure}

\begin{figure}[ht!]
    \centering
    \includegraphics[trim={2.02cm 1.15cm 2.40cm 3.04cm},clip,scale=0.335]{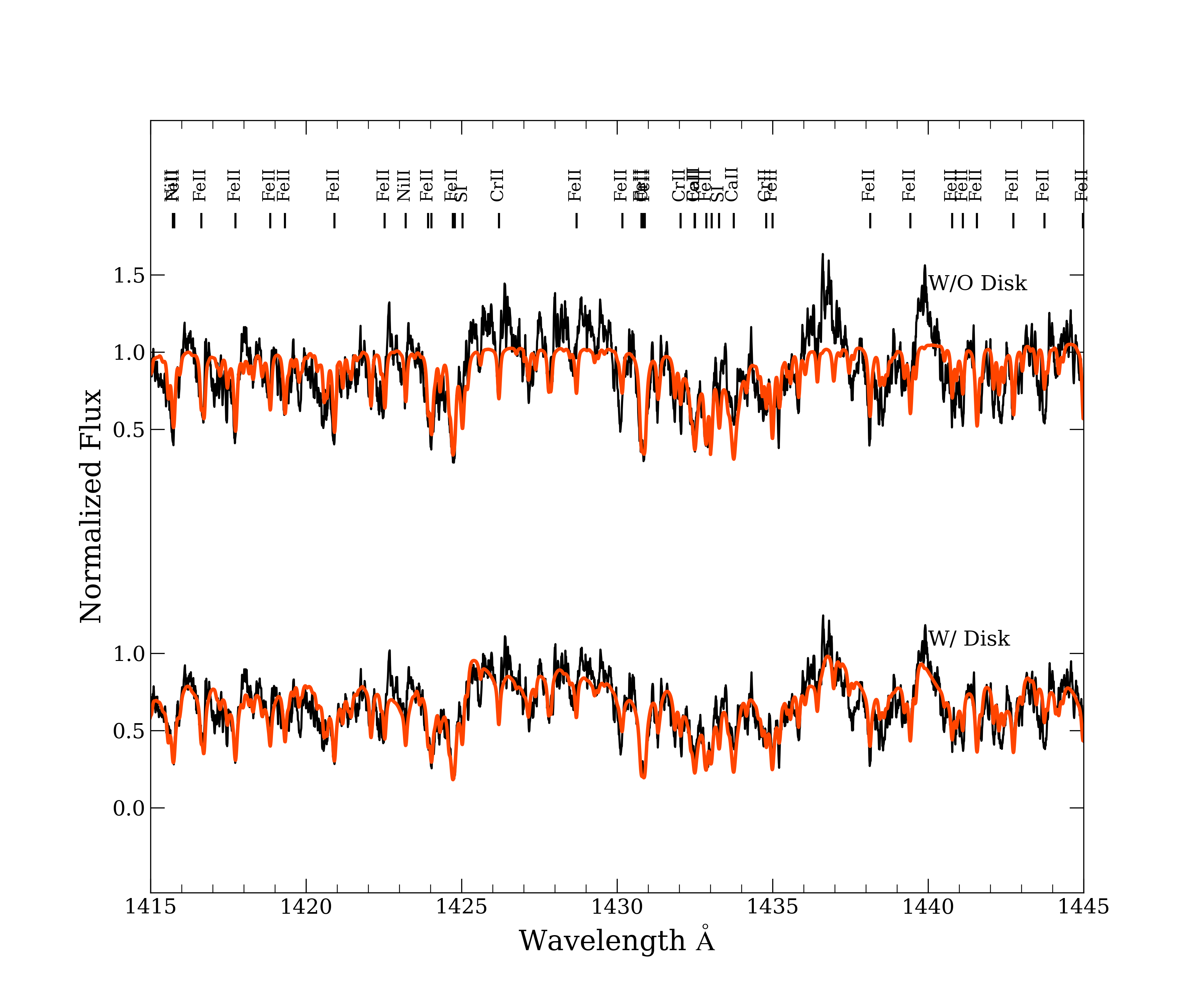}
    \caption{Same as figure \ref{fig:emissionUV02} for the 1415--1445 {\AA} region.}
    \label{fig:emissionUV02}
\end{figure}

In Figures \ref{fig:emissionUV01} and \ref{fig:emissionUV02}, we show two examples of such regions where a fit of the data without including the disk contribution appears as if there are emission lines, while those emission-like features completely disappear when we include the disk contribution. This strengthens our confidence in the ability of our disk model to correctly predict both the optical and UV absorption contributions.

\begin{figure}[h!]
\includegraphics[width=1.02\linewidth,trim={0.8cm 2.5cm 1.4cm 4.8cm},clip]{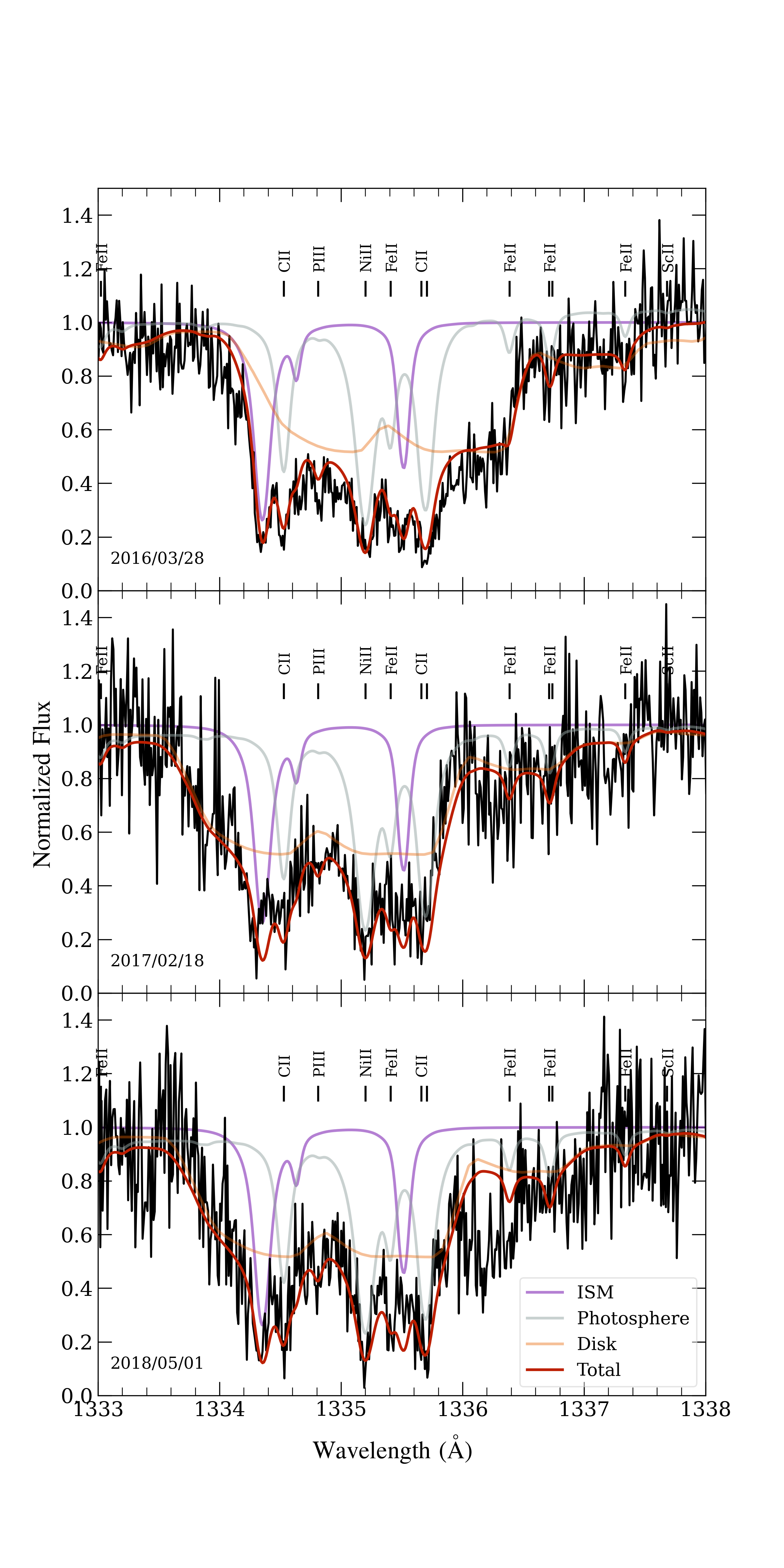}
\caption{Chosen epochs of the \ion{C}{2} doublet in the UV for the HST COS data taken on 2016 March 28, 2017 February 17 and 2018 May 1. ISM components (purple) with a velocity of -40 km/s are also included to the photospheric (grey) and disk (orange) models to fit all the features.}
\label{fig:carbon}
\end{figure} 

\begin{figure*}[ht]
        \centering
    \includegraphics[trim={3.0cm 3.9cm 3.7cm 5.9cm},clip,scale=0.56]{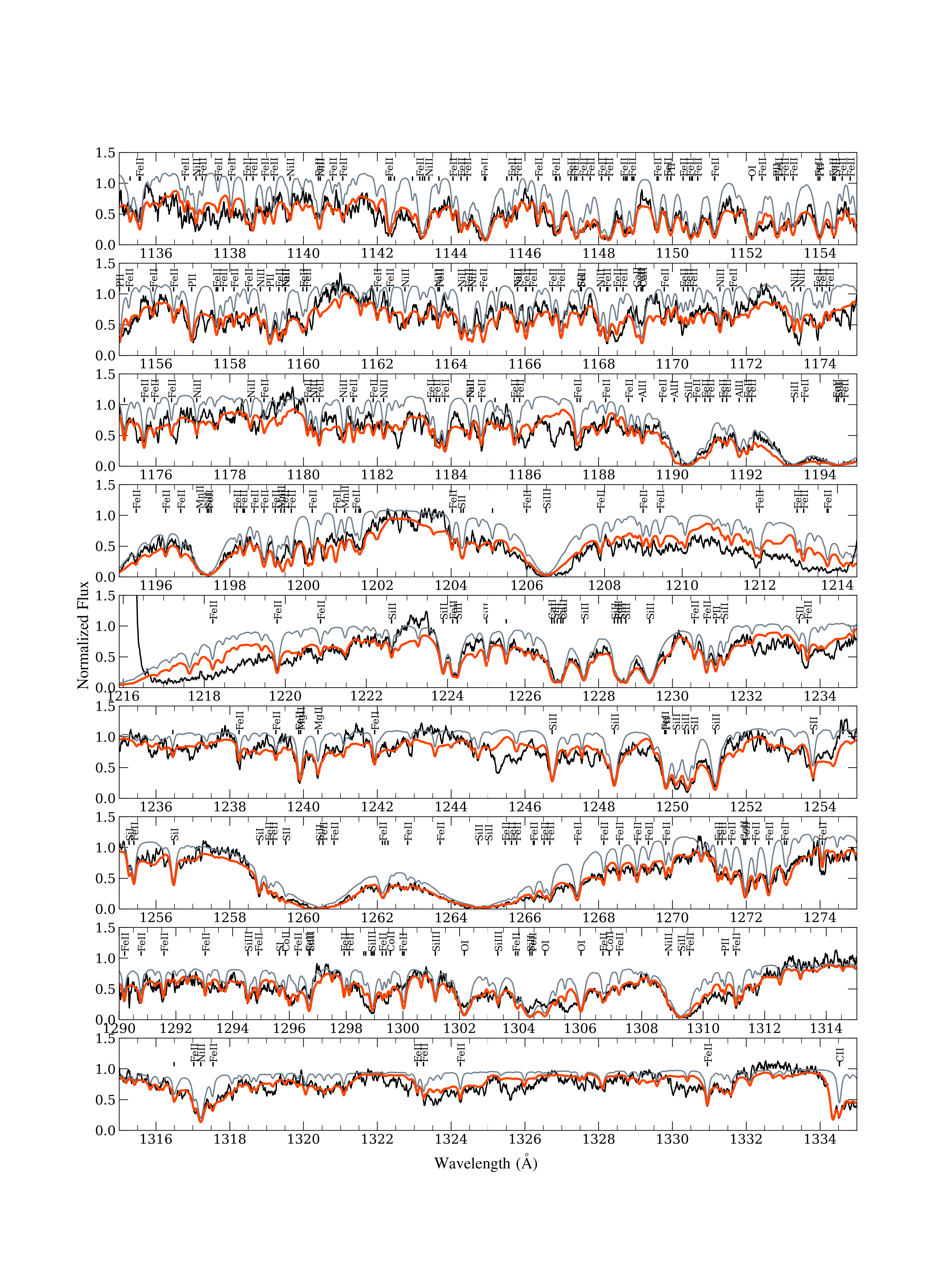}
    \caption{Selected regions of the 2016 March 28 HST COS spectrum. The contribution from the photosphere alone is in grey while the combined photospheric and circumstellar disk contributions is in orange.}
    \label{fig:20160328}
    
\end{figure*}

\begin{figure*}[ht]
        \centering
    \includegraphics[trim={3.0cm 1.7cm 3.24cm 3.4cm},clip,scale=0.56]{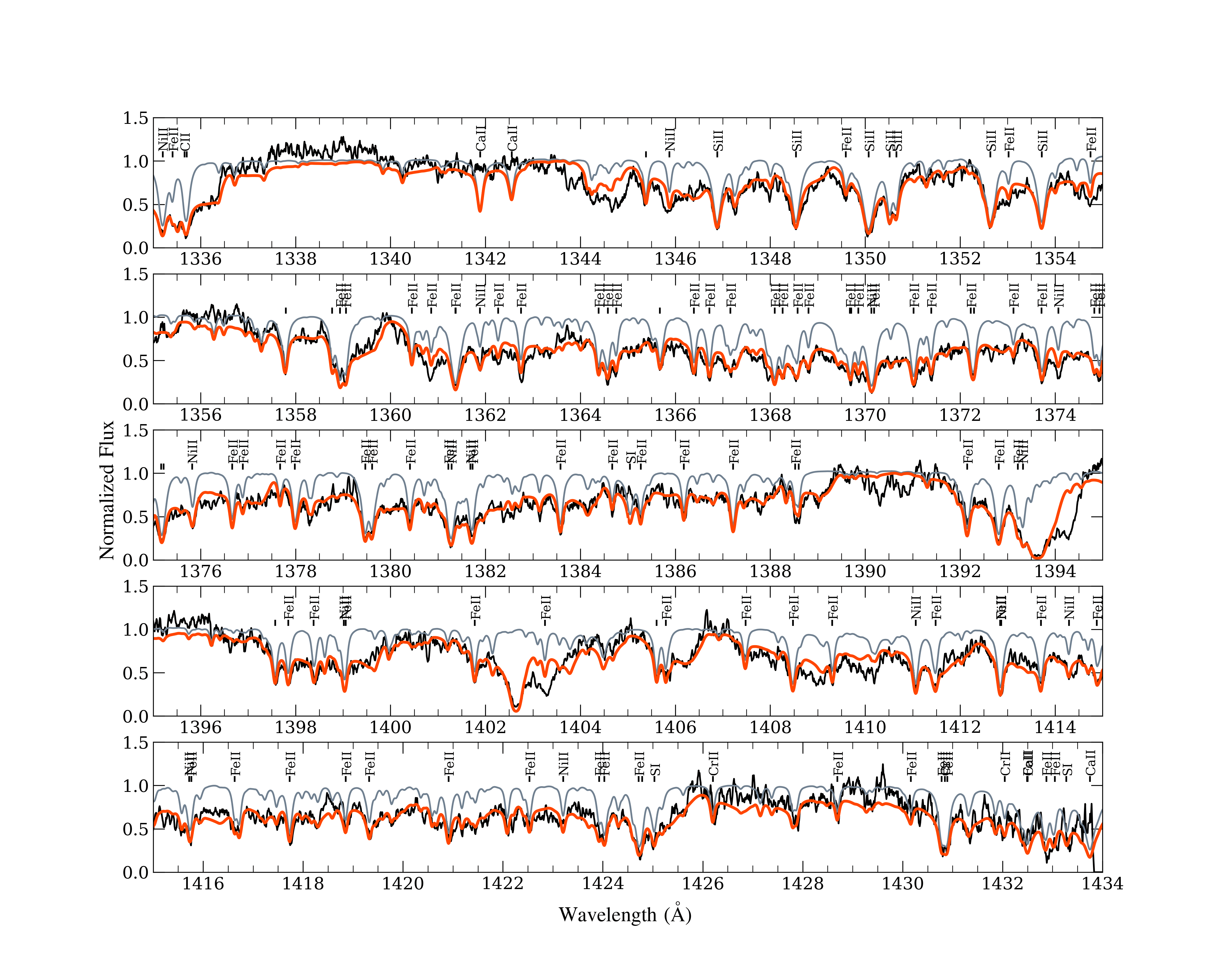}
    \subcaption{Figure \ref{fig:20160328}, continued.}
    \label{fig:20160328a}
\end{figure*}

To further highlight the importance of the disk contribution, we note that in the case of the \ion{C}{2} doublet around 1335{\,\AA}, performing the spectral analysis with the disk changed the determined abundance by almost 1~dex. In that particular case, there is also a contribution from the ISM as well as blends with multiple Fe, Ni and P lines. Figure \ref{fig:carbon} shows the relative contribution of each component for three different epochs, showing the importance of having a good representation of the disk absorption in order to distinguish correctly each contribution and derive accurate abundances for each individual epoch. For the 2018 May 1 epoch, we acknowledge that a contribution is missing in part of the 1336{\,\AA} feature. A similar phenomenon can be observed in the optical for the 2018 May 18 epoch shown in Figure \ref{fig:20180518}. As explained and showed in Figure 19 of \citet{fortin-archambault_modeling_2020}, our disk configuration is starting to fail for later epochs. Our inability to reproduce this region is thus probably related to our adopted disk's configuration for that time. It is also possible that the disk contains a denser part of the line of sight at that epoch, similar to what was observed in SDSS~J1228+1040 \citep{manser_doppler_2016}. Exploring various configurations with possibly inhomogeneous material distribution is a challenging task that is beyond the scope of this paper.

\begin{figure*}[ht!]
    \centering
    \includegraphics[trim={1.0cm 3.6cm 1.3cm 1.7cm},clip,scale=0.6]{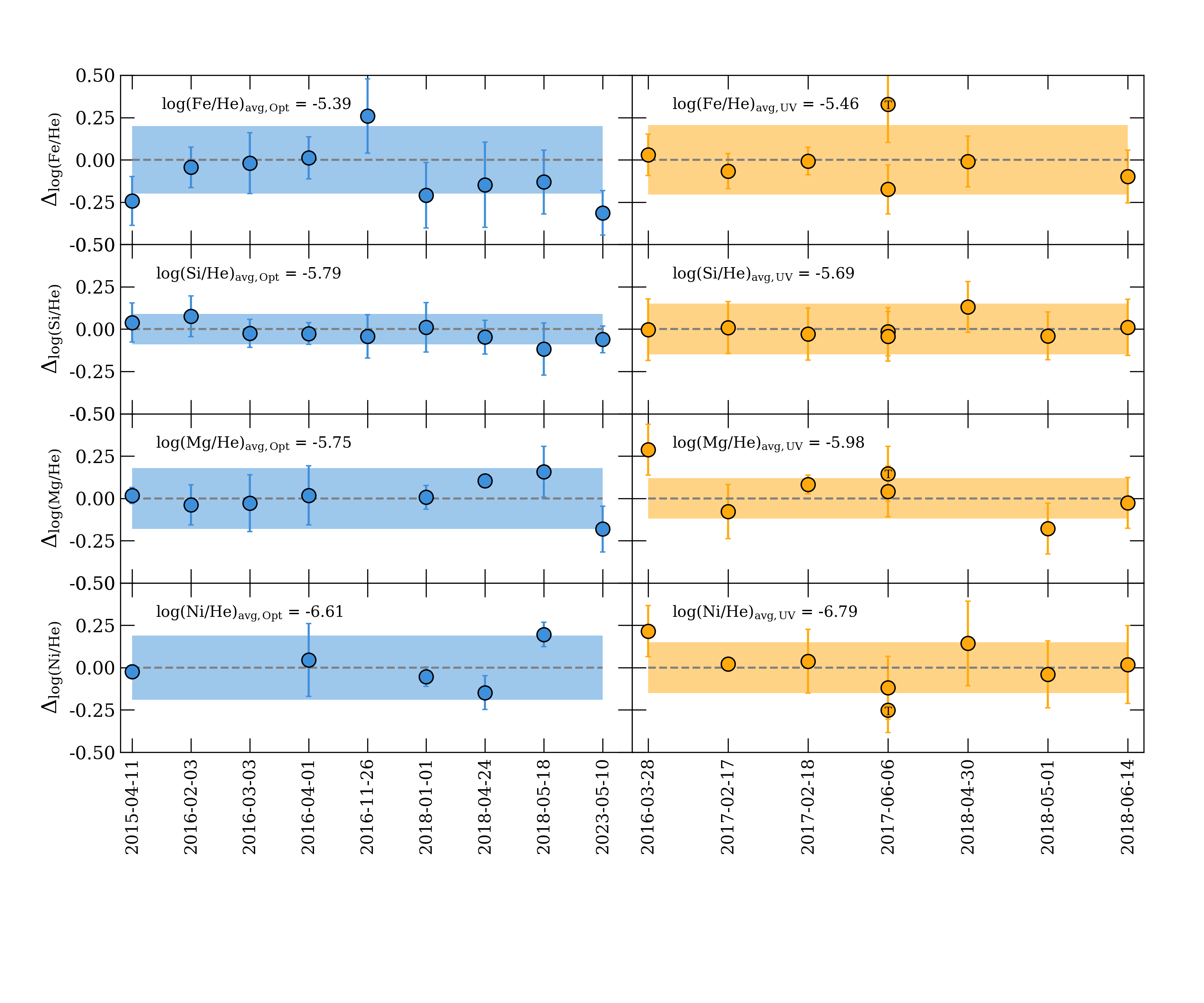}
    \caption{
    Variation of average photospheric abundance for Fe, Si, Mg, and Ni in the optical (blue) and the UV (orange) for all epochs where at least two lines were used for fitting. A \textit{T} indicates observations taken during a transit. The faded colored band represents the $1\sigma$ uncertainty (see Table \ref{tab:abn}) of the average value across all epochs.}
    \label{fig:delta_abn}
\end{figure*}

\subsection{Implication of this work on polluted WD analysis}
To our knowledge, no other metal polluted white dwarf has been observed at high resolution as frequently as WD~1145+017 over the last nine years. This unique dataset provides a rare opportunity to estimate the intrinsic uncertainties in abundance measurements. In Figure \ref{fig:delta_abn}, we show the average abundance for each epoch, both in the optical and UV, for elements whose abundances are determined from at least two spectral lines. It is interesting to note that the fluctuations, ranging from 0.1 to 0.2 dex from epoch to epoch, can be much larger than the standard uncertainty determination for each measurement.\\

While we acknowledge that our abundances for some epochs may be affected by the quality of our circumstellar feature modeling, this variation is still observed for epochs with well-reproduced circumstellar features. This suggests that chemical analysis based on only one set of data (which is almost always the case for high-resolution and/or HST observations) might systematically underestimate the uncertainties in the derived abundances.\\

We also highlight that abundances derived from spectra taken during transit (indicated by circles with a T) are severely affected. With the growing interest in similar transiting debris around white dwarfs in recent years \citep{vanderbosch_recurring_2021, aungwerojwit_long-term_2024, farihi_relentless_2022,robert_frequency_2024}, special care will be needed to avoid overinterpreting abundance patterns in such stars.

\section{Conclusion}\label{sec:conclusion}
In this work, we performed a detailed chemical analysis of the photosphere of WD~1145+017 based on spectroscopic data covering 25 epochs between 2015 and 2023. Each epoch was fitted individually, with great attention given to the inclusion of circumstellar absorption features in the fitting procedure. This necessitated major updates to the gas disk model of \citet{fortin-archambault_modeling_2020}, notably for the opacity calculations and the radial temperature profile assumed in the disk. Although the exact physical properties of the disk cannot be determined with great precision at this point due to degeneracy between its parameters, at least 10 chemical species were found to contribute to the circumstellar absorption.\\

We derived photospheric abundances for 16 different species, tying WD~1145+017 with \object{GD~362} as the most polluted white dwarf in terms of the number of heavy elements detected in its atmosphere. It was found that the composition of both the disk and the material in the photosphere is compatible with CI Chondrite composition.\\

We showed that accounting for the absorption contribution from the disk is crucial in the analysis of white dwarfs that show circumstellar features. Not properly accounting for these features in WD~1145+017 may lead to abundance errors of up to 1~dex. We also demonstrated that the circumstellar absorption forms a pseudo-continuum opacity that is dominant across the UV, causing flux reductions of up to 30\%. Hints of emission-like features observed in HST spectroscopic observations turned out to be windows with lower opacity through which the flux of the star can pass more easily. In the case of WD~1145+017, this pseudo-continuum opacity could be relatively well estimated thanks to the presence of the optical circumstellar features.  It is possible that some white dwarf stars may have gas disks that, while not opaque enough to produce optical absorption features, still reduce UV flux to a significant degree, as may be the case for \object{G29--38}, where many observations remain unexplained by previous analyses \citep{xu_elemental_2014}. If so, some previous chemical analyses based on UV observations could be affected and may need to be revisited.\\

Our analysis, based on an unprecedented number of high-resolution spectra, also suggests that uncertainties in the derived abundances could potentially be systematically underestimated in most metal-polluted white dwarf studies.\\ 

Future work should explore more complex and realistic disk configurations, as well as the effect of chemical inhomogeneity in the disk. However, since the parameter space seems to be degenerate, and performing such calculations and analyses can become quite time-consuming, implementing supervised machine learning methods will most likely be necessary to make further progress on similar systems.

\paragraph*{Acknowledgements}

{\hspace{20pt}}We thank I. Hubeny and J. Budaj for useful discussions about the use of \textsc{tlusty/synspec} opacity tables and disk characteristics.\\

 We also thank the reviewer for their insightful comments, which helped make this manuscript clearer and more accessible to scientists in related fields. ELB would like to express gratitude to L. Rogers and A. Swan for useful discussions and support throughout the writing of this paper.\\ 

This work was supported by the Natural Sciences and Engineering Research Council (NSERC) of Canada.\\ 

Based on observations obtained at the international Gemini Observatory, a program of NSF NOIRLab, which is managed by the Association of Universities for Research in Astronomy (AURA) under a cooperative agreement with the U.S. National Science Foundation on behalf of the Gemini Observatory partnership: the U.S. National Science Foundation (United States), National Research Council (Canada), Agencia Nacional de Investigaci\'{o}n y Desarrollo (Chile), Ministerio de Ciencia, Tecnolog\'{i}a e Innovaci\'{o}n (Argentina), Minist\'{e}rio da Ci\^{e}ncia, Tecnologia, Inova\c{c}\~{o}es e Comunica\c{c}\~{o}es (Brazil), and Korea Astronomy and Space Science Institute (Republic of Korea).\\ 

GHOST was built by a collaboration between Australian Astronomical Optics at Macquarie University, National Research Council Herzberg of Canada, and the Australian National University, and funded by the International Gemini partnership. The instrument scientist is Dr. Alan McConnachie at NRC, and the instrument team is also led by Dr. Gordon Robertson (at AAO), and Dr. Michael Ireland (at ANU).\\

The authors would like to acknowledge the contributions of the GHOST instrument build team, the Gemini GHOST instrument team, the full SV team, and the rest of the Gemini operations team that were involved in making the SV observations a success.

%

\facilities{Gemini:South (GHOST), HST (STIS,COS), Keck:I (HIRES), Keck:II (ESI), VLT:Kueyen (X-SHOOTER)}


\bibliography{references}

\begin{thebibliography}{}
\expandafter\ifx\csname natexlab\endcsname\relax\def\natexlab#1{#1}\fi
\providecommand{\url}[1]{\href{#1}{#1}}
\providecommand{\dodoi}[1]{doi:~\href{http://doi.org/#1}{\nolinkurl{#1}}}
\providecommand{\doeprint}[1]{\href{http://ascl.net/#1}{\nolinkurl{http://ascl.net/#1}}}
\providecommand{\doarXiv}[1]{\href{https://arxiv.org/abs/#1}{\nolinkurl{https://arxiv.org/abs/#1}}}

\bibitem[{Aungwerojwit {et~al.}(2024)Aungwerojwit, G\"{a}nsicke, Dhillon, Drake, Inight, Kaye, Marsh, Mullen, Pelisoli, \& Swan}]{aungwerojwit_long-term_2024}
Aungwerojwit, A., G\"{a}nsicke, B.~T., Dhillon, V.~S., {et~al.} 2024, Monthly Notices of the Royal Astronomical Society, 530, 117, \dodoi{10.1093/mnras/stae750}

\bibitem[{Brouwers {et~al.}(2023{\natexlab{a}})Brouwers, Bonsor, \& Malamud}]{brouwers_asynchronous_2023-1}
Brouwers, M.~G., Bonsor, A., \& Malamud, U. 2023{\natexlab{a}}, Monthly Notices of the Royal Astronomical Society, 519, 2646, \dodoi{10.1093/mnras/stac3316}

\bibitem[{Brouwers {et~al.}(2023{\natexlab{b}})Brouwers, Buchan, Bonsor, Malamud, Lynch, Rogers, \& Koester}]{brouwers_asynchronous_2023}
Brouwers, M.~G., Buchan, A.~M., Bonsor, A., {et~al.} 2023{\natexlab{b}}, Monthly Notices of the Royal Astronomical Society, 519, 2663, \dodoi{10.1093/mnras/stac3317}

\bibitem[{Budaj {et~al.}(2022)Budaj, Maliuk, \& Hubeny}]{budaj_wd_2022}
Budaj, J., Maliuk, A., \& Hubeny, I. 2022, Astronomy \& Astrophysics, 660, A72, \dodoi{10.1051/0004-6361/202141924}

\bibitem[{Capitanio {et~al.}(2017)Capitanio, Lallement, Vergely, Elyajouri, \& Monreal-Ibero}]{capitanio_three-dimensional_2017}
Capitanio, L., Lallement, R., Vergely, J.~L., Elyajouri, M., \& Monreal-Ibero, A. 2017, Astronomy and Astrophysics, 606, A65, \dodoi{10.1051/0004-6361/201730831}

\bibitem[{Cauley {et~al.}(2018)Cauley, Farihi, Redfield, Bachman, Parsons, \& G\"{a}nsicke}]{cauley_evidence_2018}
Cauley, P.~W., Farihi, J., Redfield, S., {et~al.} 2018, The Astrophysical Journal, 852, L22, \dodoi{10.3847/2041-8213/aaa3d9}

\bibitem[{Coutu {et~al.}(2019)Coutu, Dufour, Bergeron, Blouin, Loranger, Allard, \& Dunlap}]{coutu_analysis_2019}
Coutu, S., Dufour, P., Bergeron, P., {et~al.} 2019, The Astrophysical Journal, 885, 74, \dodoi{10.3847/1538-4357/ab46b9}

\bibitem[{Debes \& Sigurdsson(2002)}]{debes_are_2002}
Debes, J.~H., \& Sigurdsson, S. 2002, The Astrophysical Journal, 572, 556, \dodoi{10.1086/340291}

\bibitem[{Dufour {et~al.}(2016)Dufour, Blouin, Coutu, Fortin-Archambault, Thibeault, Bergeron, \& Fontaine}]{dufour_montreal_2016}
Dufour, P., Blouin, S., Coutu, S., {et~al.} 2016, \dodoi{10.48550/ARXIV.1610.00986}

\bibitem[{Dufour {et~al.}(2012)Dufour, Kilic, Fontaine, Bergeron, Melis, \& Bochanski}]{dufour_detailed_2012}
Dufour, P., Kilic, M., Fontaine, G., {et~al.} 2012, The Astrophysical Journal, 749, 6, \dodoi{10.1088/0004-637X/749/1/6}

\bibitem[{Farihi {et~al.}(2022)Farihi, Hermes, Marsh, Mustill, Wyatt, Guidry, Wilson, Redfield, Izquierdo, Toloza, G\"{a}nsicke, Aungwerojwit, Kaewmanee, Dhillon, \& Swan}]{farihi_relentless_2022}
Farihi, J., Hermes, J.~J., Marsh, T.~R., {et~al.} 2022, Monthly Notices of the Royal Astronomical Society, 511, 1647, \dodoi{10.1093/mnras/stab3475}

\bibitem[{Fortin-Archambault {et~al.}(2020)Fortin-Archambault, Dufour, \& Xu}]{fortin-archambault_modeling_2020}
Fortin-Archambault, M., Dufour, P., \& Xu, S. 2020, The Astrophysical Journal, 888, 47, \dodoi{10.3847/1538-4357/ab585a}

\bibitem[{G\"{a}nsicke {et~al.}(2012)G\"{a}nsicke, Koester, Farihi, Girven, Parsons, \& Breedt}]{gansicke_chemical_2012}
G\"{a}nsicke, B.~T., Koester, D., Farihi, J., {et~al.} 2012, Monthly Notices of the Royal Astronomical Society, 424, 333, \dodoi{10.1111/j.1365-2966.2012.21201.x}

\bibitem[{Hubeny \& Lanz(2017)}]{hubeny_brief_2017}
Hubeny, I., \& Lanz, T. 2017, A brief introductory guide to {TLUSTY} and {SYNSPEC}, \dodoi{10.48550/arXiv.1706.01859}

\bibitem[{Hubeny {et~al.}(2021)Hubeny, Prieto, Osorio, \& Lanz}]{hubeny_tlusty_2021}
Hubeny, I., Prieto, C.~A., Osorio, Y., \& Lanz, T. 2021, \dodoi{10.48550/ARXIV.2104.02829}

\bibitem[{Jessberger {et~al.}(1988)Jessberger, Christoforidis, \& Kissel}]{jessberger_aspects_1988}
Jessberger, E.~K., Christoforidis, A., \& Kissel, J. 1988, Nature, 332, 691, \dodoi{10.1038/332691a0}

\bibitem[{Jura(2008)}]{jura_pollution_2008}
Jura, M. 2008, The Astronomical Journal, 135, 1785, \dodoi{10.1088/0004-6256/135/5/1785}

\bibitem[{Jura {et~al.}(2012)Jura, Xu, Klein, Koester, \& Zuckerman}]{jura_two_2012}
Jura, M., Xu, S., Klein, B., Koester, D., \& Zuckerman, B. 2012, The Astrophysical Journal, 750, 69, \dodoi{10.1088/0004-637X/750/1/69}

\bibitem[{Jura \& Young(2014)}]{jura_extrasolar_2014}
Jura, M., \& Young, E.~D. 2014, Annual Review of Earth and Planetary Sciences, 42, 45, \dodoi{10.1146/annurev-earth-060313-054740}

\bibitem[{Klein {et~al.}(2010)Klein, Jura, Koester, Zuckerman, \& Melis}]{klein_chemical_2010}
Klein, B., Jura, M., Koester, D., Zuckerman, B., \& Melis, C. 2010, The Astrophysical Journal, 709, 950, \dodoi{10.1088/0004-637X/709/2/950}

\bibitem[{Klein {et~al.}(2021)Klein, Doyle, Zuckerman, Dufour, Blouin, Melis, Weinberger, \& Young}]{klein_discovery_2021}
Klein, B.~L., Doyle, A.~E., Zuckerman, B., {et~al.} 2021, The Astrophysical Journal, 914, 61, \dodoi{10.3847/1538-4357/abe40b}

\bibitem[{Koester(2009)}]{koester_accretion_2009}
Koester, D. 2009, Astronomy and Astrophysics, 498, 517, \dodoi{10.1051/0004-6361/200811468}

\bibitem[{Koester {et~al.}(2014)Koester, G\"{a}nsicke, \& Farihi}]{koester_frequency_2014}
Koester, D., G\"{a}nsicke, B.~T., \& Farihi, J. 2014, Astronomy and Astrophysics, 566, A34, \dodoi{10.1051/0004-6361/201423691}

\bibitem[{Labrie {et~al.}(2023)Labrie, Simpson, Cardenes, Turner, Soraisam, Quint, Oberdorf, Placco, Berke, Smirnova, Conseil, Vacca, \& Thomas-Osip}]{labrie_dragons-quick_2023}
Labrie, K., Simpson, C., Cardenes, R., {et~al.} 2023, Research Notes of the American Astronomical Society, 7, 214, \dodoi{10.3847/2515-5172/ad0044}

\bibitem[{Lodders(2003)}]{lodders_solar_2003}
Lodders, K. 2003, The Astrophysical Journal, 591, 1220, \dodoi{10.1086/375492}

\bibitem[{Manser {et~al.}(2016)Manser, G\"{a}nsicke, Marsh, Veras, Koester, Breedt, Pala, Parsons, \& Southworth}]{manser_doppler_2016}
Manser, C.~J., G\"{a}nsicke, B.~T., Marsh, T.~R., {et~al.} 2016, Monthly Notices of the Royal Astronomical Society, 455, 4467, \dodoi{10.1093/mnras/stv2603}

\bibitem[{Manser {et~al.}(2024)Manser, G\"{a}nsicke, Izquierdo, Swan, Najita, Rockosi, Carrillo, Kim, Xu, Dey, Aguilar, Ahlen, Blum, Brooks, Claybaugh, Dawson, de~la Macorra, Doel, Gaztañaga, Gontcho A~Gontcho, Honscheid, Kehoe, Kremin, Landriau, Le~Guillou, Levi, Li, Meisner, Miquel, Nie, Rezaie, Rossi, Sanchez, Schubnell, Tarl\'{e}, Weaver, Zhou, \& Zou}]{manser_frequency_2024}
Manser, C.~J., G\"{a}nsicke, B.~T., Izquierdo, P., {et~al.} 2024, Monthly Notices of the Royal Astronomical Society, 531, L27, \dodoi{10.1093/mnrasl/slae026}

\bibitem[{McConnachie {et~al.}(2024)McConnachie, Hayes, Robertson, Pazder, Ireland, Burley, Churilov, Lothrop, Zhelem, Kalari, Anthony, Baker, Berg, Chapin, Chin, Densmore, Diaz, Dunn, Edgar, Farrell, Firpo, Fuentes, Gomez-Jimenez, Hardy, Henderson, Hill, Labrie, Jensen, Lambert, Lawrence, Macdonald, Margheim, Millar, Muller, Nielsen, P\'{e}rez, Quiroz, Ruiz-Carmona, Sebo, Sestito, Silva, Simpson, Smith, Venkatesan, Waller, Waller, Wevers, Venn, Young, \& Silversides}]{mcconnachie_science_2024}
McConnachie, A.~W., Hayes, C.~R., Robertson, J.~G., {et~al.} 2024, Publications of the Astronomical Society of the Pacific, 136, 035001, \dodoi{10.1088/1538-3873/ad1ed4}

\bibitem[{McDonough(2003)}]{mcdonough_compositional_2003}
McDonough, W. 2003, in Treatise on {Geochemistry} (Elsevier), 547--568, \dodoi{10.1016/B0-08-043751-6/02015-6}

\bibitem[{Melis \& Dufour(2017)}]{melis_does_2017}
Melis, C., \& Dufour, P. 2017, The Astrophysical Journal, 834, 1, \dodoi{10.3847/1538-4357/834/1/1}

\bibitem[{Paquette {et~al.}(1986)Paquette, Pelletier, Fontaine, \& Michaud}]{paquette_diffusion_1986}
Paquette, C., Pelletier, C., Fontaine, G., \& Michaud, G. 1986, The Astrophysical Journal Supplement Series, 61, 197, \dodoi{10.1086/191112}

\bibitem[{Pringle(1981)}]{pringle_accretion_1981}
Pringle, J.~E. 1981, Annual Review of Astronomy and Astrophysics, 19, 137, \dodoi{10.1146/annurev.aa.19.090181.001033}

\bibitem[{Putirka \& Xu(2021)}]{putirka_polluted_2021}
Putirka, K.~D., \& Xu, S. 2021, Nature Communications, 12, 6168, \dodoi{10.1038/s41467-021-26403-8}

\bibitem[{Rappaport {et~al.}(2016)Rappaport, Gary, Kaye, Vanderburg, Croll, Benni, \& Foote}]{rappaport_drifting_2016}
Rappaport, S., Gary, B.~L., Kaye, T., {et~al.} 2016, Monthly Notices of the Royal Astronomical Society, 458, 3904, \dodoi{10.1093/mnras/stw612}

\bibitem[{Robert {et~al.}(2024)Robert, Farihi, Van~Eylen, Aungwerojwit, G\"{a}nsicke, Redfield, Dhillon, Marsh, \& Swan}]{robert_frequency_2024}
Robert, A., Farihi, J., Van~Eylen, V., {et~al.} 2024, Monthly Notices of the Royal Astronomical Society, \dodoi{10.1093/mnras/stae1859}

\bibitem[{Rogers {et~al.}(2024{\natexlab{a}})Rogers, Bonsor, Xu, Dufour, Klein, Buchan, Hodgkin, Hardy, Kissler-Patig, Melis, Weinberger, \& Zuckerman}]{rogers_seven_2024}
Rogers, L.~K., Bonsor, A., Xu, S., {et~al.} 2024{\natexlab{a}}, Monthly Notices of the Royal Astronomical Society, 527, 6038, \dodoi{10.1093/mnras/stad3557}

\bibitem[{Rogers {et~al.}(2024{\natexlab{b}})Rogers, Bonsor, Xu, Buchan, Dufour, Klein, Hodgkin, Kissler-Patig, Melis, Walton, \& Weinberger}]{rogers_seven_2024-1}
---. 2024{\natexlab{b}}, Monthly Notices of the Royal Astronomical Society, \dodoi{10.1093/mnras/stae1520}

\bibitem[{Swan {et~al.}(2019)Swan, Farihi, Koester, Hollands, Parsons, Cauley, Redfield, \& G\"{a}nsicke}]{swan_interpretation_2019}
Swan, A., Farihi, J., Koester, D., {et~al.} 2019, Monthly Notices of the Royal Astronomical Society, 490, 202, \dodoi{10.1093/mnras/stz2337}

\bibitem[{Swan {et~al.}(2023)Swan, Farihi, Melis, Dufour, Desch, Koester, \& Guo}]{swan_planetesimals_2023}
Swan, A., Farihi, J., Melis, C., {et~al.} 2023, Monthly Notices of the Royal Astronomical Society, stad2867, \dodoi{10.1093/mnras/stad2867}

\bibitem[{Van~Maanen(1917)}]{van_maanen_two_1917}
Van~Maanen, A. 1917, Publications of the Astronomical Society of the Pacific, 29, 258, \dodoi{10.1086/122654}

\bibitem[{Vanderbosch {et~al.}(2021)Vanderbosch, Rappaport, Guidry, Gary, Blouin, Kaye, Weinberger, Melis, Klein, Zuckerman, Vanderburg, Hermes, Hegedus, Burleigh, Sefako, Worters, \& Heintz}]{vanderbosch_recurring_2021}
Vanderbosch, Z.~P., Rappaport, S., Guidry, J.~A., {et~al.} 2021, The Astrophysical Journal, 917, 41, \dodoi{10.3847/1538-4357/ac0822}

\bibitem[{Vanderburg {et~al.}(2015)Vanderburg, Johnson, Rappaport, Bieryla, Irwin, Lewis, Kipping, Brown, Dufour, Ciardi, Angus, Schaefer, Latham, Charbonneau, Beichman, Eastman, McCrady, Wittenmyer, \& Wright}]{vanderburg_disintegrating_2015}
Vanderburg, A., Johnson, J.~A., Rappaport, S., {et~al.} 2015, Nature, 526, 546, \dodoi{10.1038/nature15527}

\bibitem[{Wilson {et~al.}(2019)Wilson, Farihi, G\"{a}nsicke, \& Swan}]{wilson_unbiased_2019}
Wilson, T.~G., Farihi, J., G\"{a}nsicke, B.~T., \& Swan, A. 2019, Monthly Notices of the Royal Astronomical Society, 487, 133, \dodoi{10.1093/mnras/stz1050}

\bibitem[{Xu {et~al.}(2019{\natexlab{a}})Xu, Dufour, Klein, Melis, Monson, Zuckerman, Young, \& Jura}]{xu_compositions_2019}
Xu, S., Dufour, P., Klein, B., {et~al.} 2019{\natexlab{a}}, The Astronomical Journal, 158, 242, \dodoi{10.3847/1538-3881/ab4cee}

\bibitem[{Xu {et~al.}(2016)Xu, Jura, Dufour, \& Zuckerman}]{xu_evidence_2016}
Xu, S., Jura, M., Dufour, P., \& Zuckerman, B. 2016, The Astrophysical Journal, 816, L22, \dodoi{10.3847/2041-8205/816/2/L22}

\bibitem[{Xu {et~al.}(2014)Xu, Jura, Koester, Klein, \& Zuckerman}]{xu_elemental_2014}
Xu, S., Jura, M., Koester, D., Klein, B., \& Zuckerman, B. 2014, The Astrophysical Journal, 783, 79, \dodoi{10.1088/0004-637X/783/2/79}

\bibitem[{Xu {et~al.}(2019{\natexlab{b}})Xu, Hallakoun, Gary, Dalba, Debes, Dufour, Fortin-Archambault, Fukui, Jura, Klein, Kusakabe, Muirhead, Narita, Steele, Su, Vanderburg, Watanabe, Zhan, \& Zuckerman}]{xu_shallow_2019}
Xu, S., Hallakoun, N., Gary, B., {et~al.} 2019{\natexlab{b}}, The Astronomical Journal, 157, 255, \dodoi{10.3847/1538-3881/ab1b36}

\bibitem[{Zuckerman {et~al.}(2007)Zuckerman, Koester, Melis, Hansen, \& Jura}]{zuckerman_chemical_2007}
Zuckerman, B., Koester, D., Melis, C., Hansen, B.~M., \& Jura, M. 2007, The Astrophysical Journal, 671, 872, \dodoi{10.1086/522223}

\bibitem[{Zuckerman {et~al.}(2003)Zuckerman, Koester, Reid, \& H\"{u}nsch}]{zuckerman_metal_2003}
Zuckerman, B., Koester, D., Reid, I.~N., \& H\"{u}nsch, M. 2003, The Astrophysical Journal, 596, 477, \dodoi{10.1086/377492}

\end{thebibliography}
\bibliographystyle{aasjournal}

\newpage
\vspace{-24pt}
\appendix
\vspace{-18pt}
\section{Plots of selected epochs}\label{apx:fit_epochs}
\vspace{-10pt}
\begin{figure*}[h]
        \centering
    \includegraphics[trim={3.2cm 3.9cm 3.4cm 5.9cm},clip,scale=0.49]{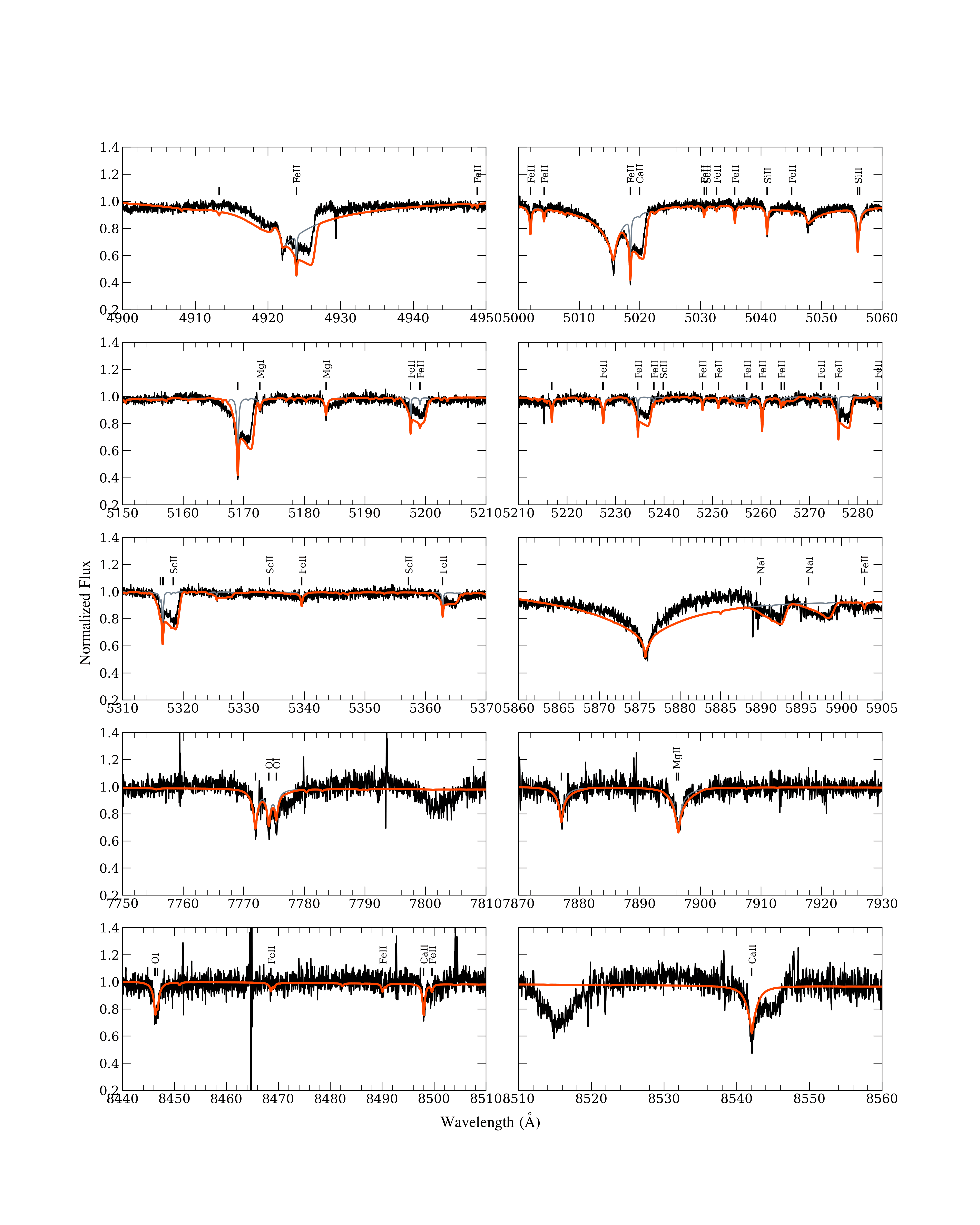}
    \caption{Selected regions of the 2016 February 3 HIRES spectrum. The contribution from the photosphere alone is in grey while the combined photospheric and circumstellar disk contributions is in orange.}
    \label{fig:duf12_20160203-1}
\end{figure*}

\begin{figure*}[h]
    \centering
    \includegraphics[trim={3.20cm 3.9cm 3.4cm 5.9cm},clip,scale=0.5]{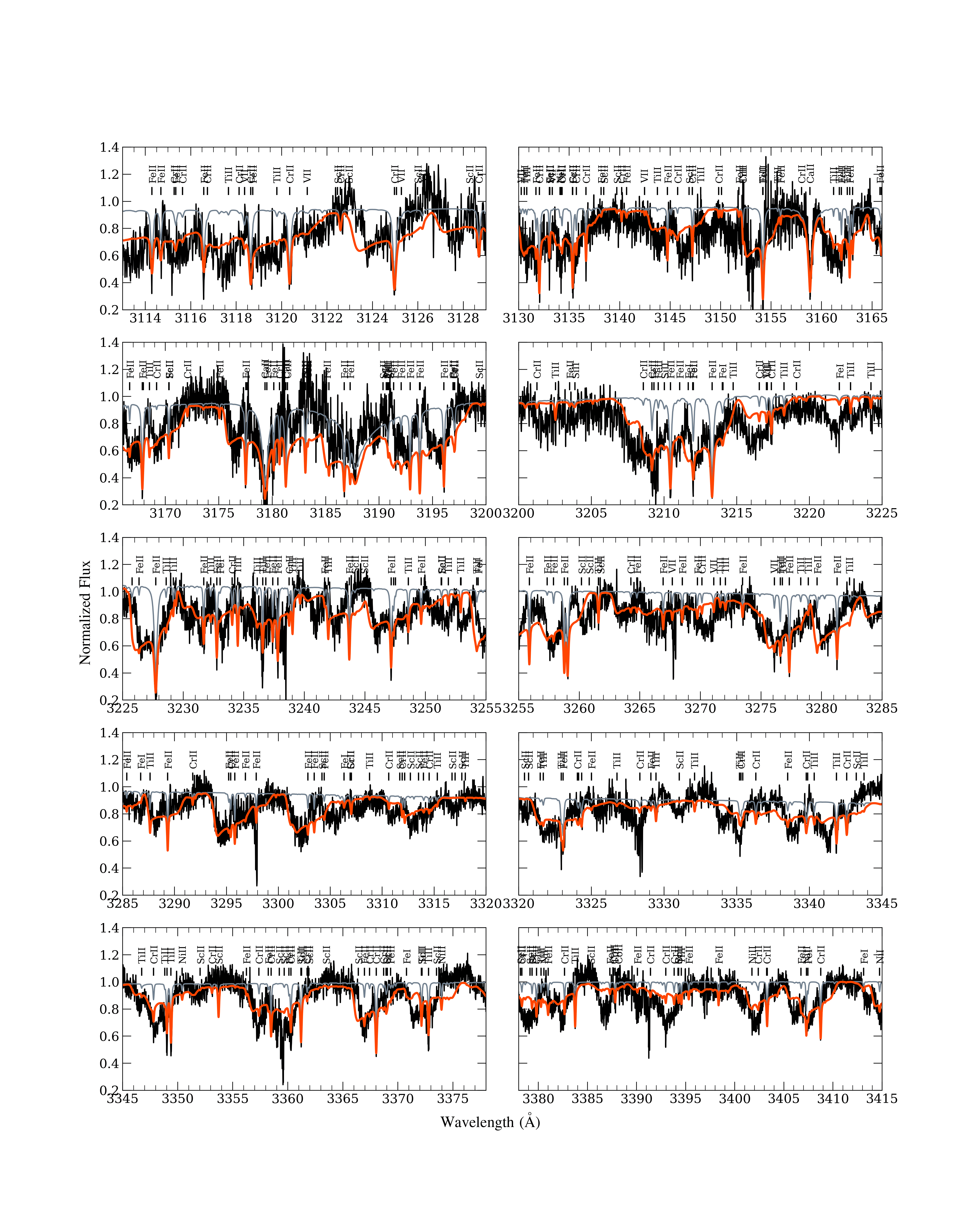}
    \caption{Selected regions of the 2018 May 18 HIRES spectrum. The contribution from the photosphere alone is in grey while the combined photospheric and circumstellar disk contributions is in orange.}
    \label{fig:20180518}
    
\end{figure*}

\begin{figure*}[ht]
        \centering
    \includegraphics[trim={3.20cm 3.9cm 3.4cm 5.9cm},clip,scale=0.5]{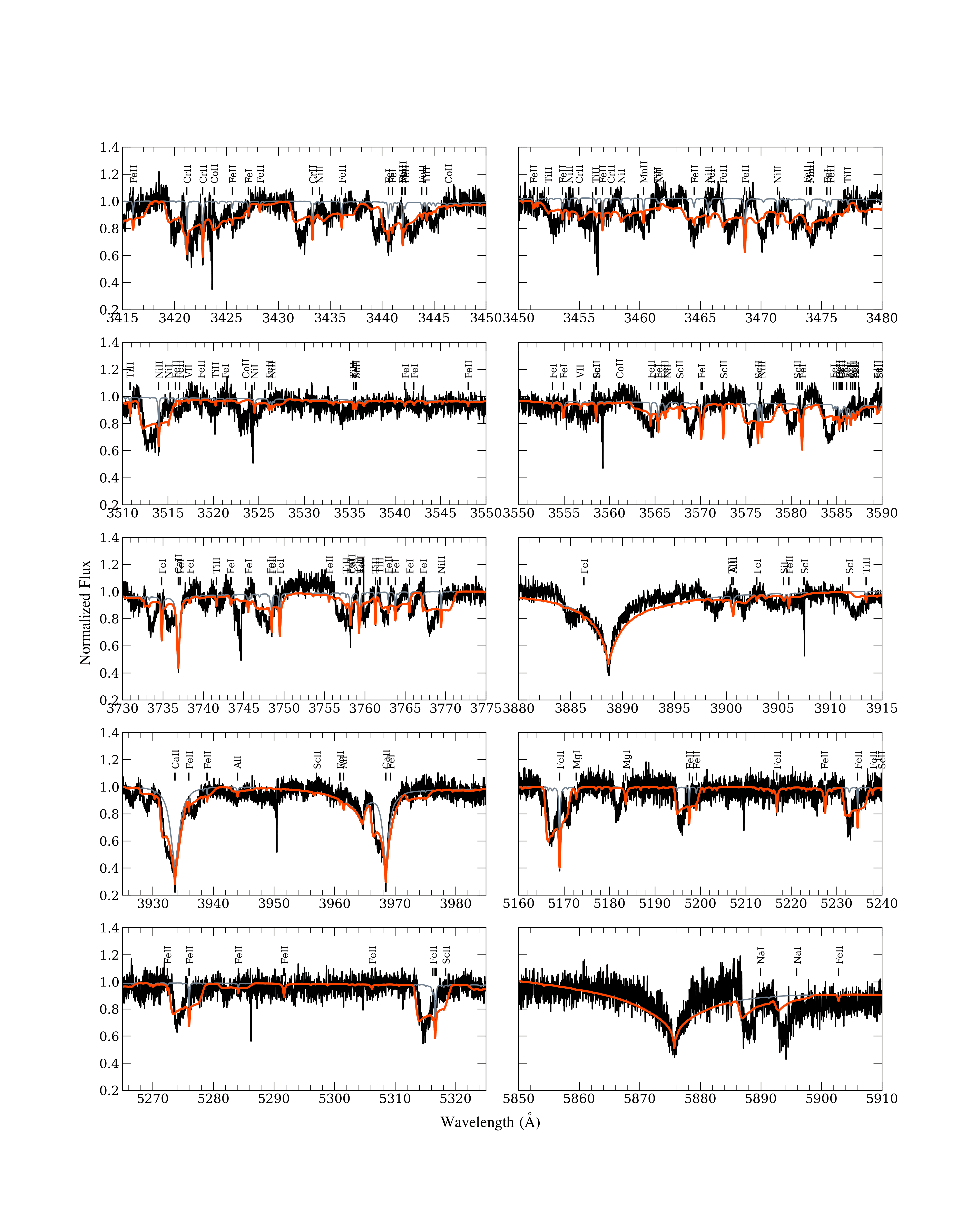}
    \subcaption{Figure \ref{fig:20180518}, continued.}
    \label{fig:20180518a}
\end{figure*}

\begin{figure*}[h!]
        \centering
    \includegraphics[trim={3.3cm 3.9cm 3.4cm 5.9cm},clip,scale=0.5]{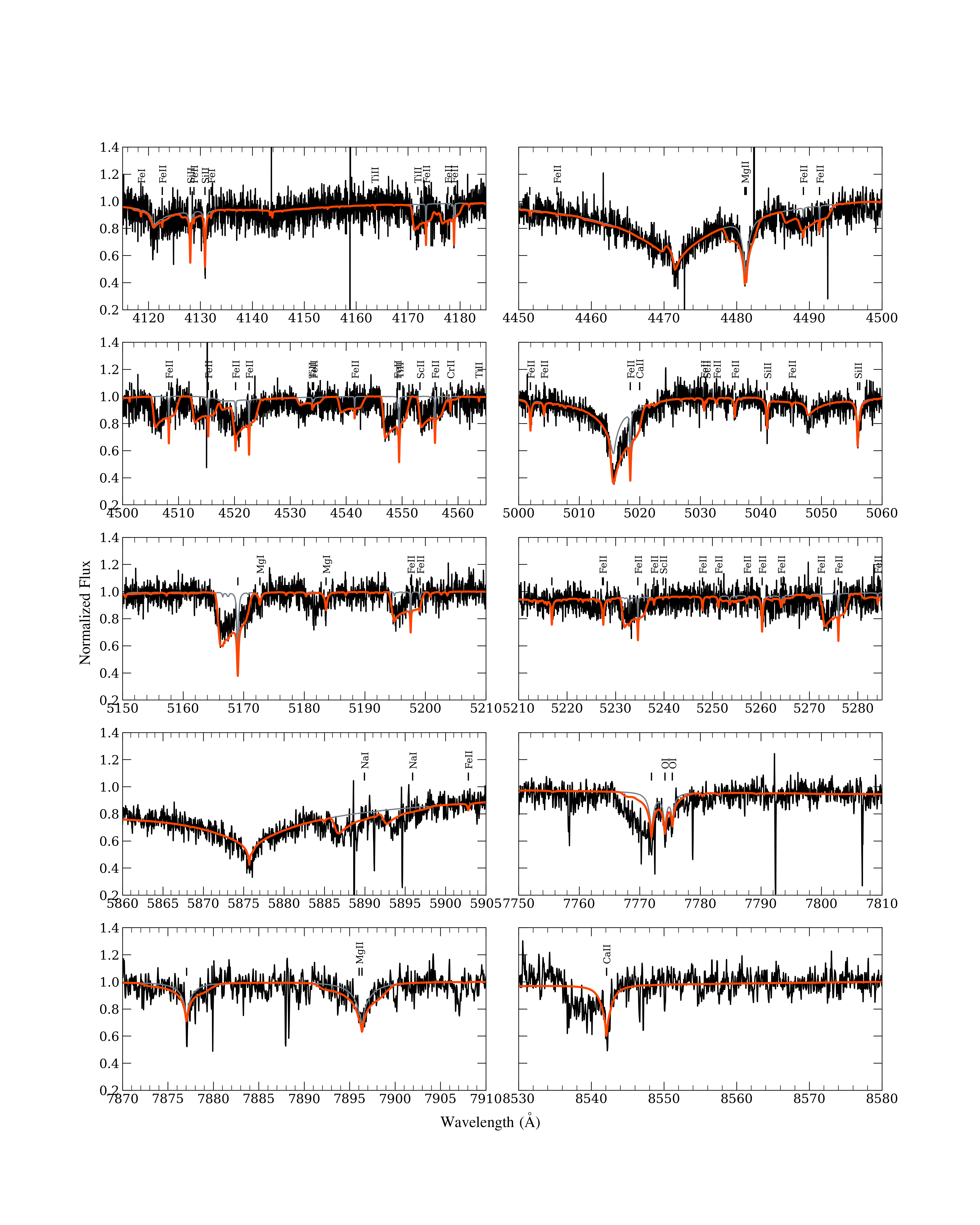}
    \caption{Selected regions of the 2023 May 10 GHOST spectrum. The contribution from the photosphere alone is in grey while the combined photospheric and circumstellar disk contributions is in orange.}
    \label{fig:duf12-2023-1}
\end{figure*}

\end{document}